\documentclass{jfm}

\usepackage{graphicx}
\usepackage{newtxtext}
\usepackage{newtxmath}
\usepackage{ragged2e}
\usepackage{blindtext}
\usepackage{natbib}
\usepackage{xcolor}
\usepackage{hyperref}

\hypersetup{
colorlinks = true,
urlcolor   = blue,
citecolor  = blue,
}
\DeclareUnicodeCharacter{2009}{\,} 

\usepackage{caption}

\newcommand{\RomanNumeralCaps}[1]
\linenumbers
\usepackage[format=plain]{caption}

\makeatletter
\patchcmd{\NAT@citex}
  {\@citea\NAT@hyper@{%
     \NAT@nmfmt{\NAT@nm}%
     \hyper@natlinkbreak{\NAT@aysep\NAT@spacechar}{\@citeb\@extra@b@citeb}%
     \NAT@date}}
  {\@citea\NAT@nmfmt{\NAT@nm}%
   \NAT@aysep\NAT@spacechar\NAT@hyper@{\NAT@date}}{}{}

\patchcmd{\NAT@citex}
  {\@citea\NAT@hyper@{%
     \NAT@nmfmt{\NAT@nm}%
     \hyper@natlinkbreak{\NAT@spacechar\NAT@@open\if*#1*\else#1\NAT@spacechar\fi}%
       {\@citeb\@extra@b@citeb}%
     \NAT@date}}
  {\@citea\NAT@nmfmt{\NAT@nm}%
   \NAT@spacechar\NAT@@open\if*#1*\else#1\NAT@spacechar\fi\NAT@hyper@{\NAT@date}}
  {}{}

\makeatother

\title{Fluid viscoelasticity controls acoustic streaming via shear waves}

\author{
  T. Sujith\aff{1}
 \and A.K. Sen\aff{1} \corresp{\email{ashis@iitm.ac.in}}}

\affiliation{\aff{1}Micro Nano Bio Fluidics Unit, Department of Mechanical Engineering,
Indian Institute of Technology Madras, Chennai -- 600036, Tamil Nadu, India
}

\begin{document}
\maketitle

\begin{abstract}
Control of acoustic streaming can significantly impact fluid and particle transport in microfluidics. We report enhancement, suppression, and reversal of acoustic streaming inside a rectangular microchannel by controlling the fluid viscoelastic properties. Our study reveals that the streaming regimes depend on Deborah number ($De$) and viscous diffusion number ($Dv$), expressed in terms of a Streaming Coefficient ($C_s$). We find streaming is enhanced when $C_s>1$, suppressed for $0\leq C_s\leq1$, and reversed when $C_s<0$. We explain the regimes in terms of the interplay between the Reynolds and viscoelastic stresses that collectively drive fluid motion. Remarkably, we discover the role of viscoelastic shear waves in acoustic streaming transition characterized by the ratio of acoustic attenuation length and shear wavelength. We gain deeper insight into the streaming transition by examining energy dynamics in terms of the loss and storage moduli. Our study may find applications in acousto-microfluidics systems for particle handling and fluid pumping/mixing.
\end{abstract}


\section{Introduction}
\label{sec:headings}

 Oscillatory motion in a fluid, driven by acoustic waves or vibrating boundaries, produces a steady flow known as acoustic streaming \citep{Faraday1831,LIGHTHILL1978,Eckart1948,Schlichting1932,Rayleigh1884}. This nonlinear effect stems from the dissipation of acoustic energy \citep{LIGHTHILL1978} and arises through two primary mechanisms. In large-scale systems, where dimensions far exceed the acoustic wavelength, bulk attenuation of traveling waves leads to “quartz wind” \citep{Eckart1948}. In contrast, in small-scale systems, energy dissipation within a thin boundary layer gives rise to “boundary layer-driven streaming” \citep{Schlichting1932, Rayleigh1884}. The vorticity generated within the boundary layer, known as “inner streaming” or “Schlichting streaming” \citep{Schlichting1932}, drives counter-rotating vortices in the bulk fluid, referred to as “outer streaming” or “Rayleigh streaming” \citep{Rayleigh1884}. Boundary-layer-driven streaming has gained attention in acousto-microfluidic systems for applications in pumping, mixing, and particle/cell manipulation \citep{Wiklund2012,Sadhal2012}. In such systems, acoustic waves at MHz frequencies generate acoustic radiation forces (ARF) \citep{Bruus2012} in addition to streaming effects \citep{Muller2012}. ARF align suspended particles while streaming, governed by the Stokes drag, tend to disorganize them, particularly at the submicron scale \citep{Bruus2012, Muller2012}. Suppression of streaming can greatly benefit the focusing and sorting of submicron particles such as viruses, bacteria, and exosomes for medical diagnostics \citep{VanAssche2020}. On the other hand, enhanced streaming is crucial for overcoming low Reynolds number constraints to improve fluid mixing and pumping \citep{Wiklund2012}. Thus, achieving effective streaming control is critical in acousto-microfluidic systems. Existing streaming suppression methods rely on fluid inhomogeneity \citep{Karlsen2018} or shape-optimized channels \citep{Bach2020}, while streaming enhancement has been achieved using thermal gradients \citep{Qiu2021}; however, these strategies are difficult to implement and offer limited control. Currently, no method enables both suppression and enhancement of acoustic streaming in single-fluid microchannel systems, and none enables its directional reversal. This underscores the need for a simple and effective alternative approach to address a critical gap in the field. Prior studies have examined acoustic streaming in viscoelastic (VE) fluids around a particle \citep{Doinikov2021a} and in a narrow channel of depth comparable to the boundary layer thickness \citep{Vargas2022}. In this study, we report the enhancement, suppression, and reversal of acoustic streaming in a typical rectangular microchannel by tuning the fluid viscoelasticity.  More importantly, we take a profound approach to elucidate the physics underlying acoustic streaming control in terms of relevant timescales, stresses, and viscoelastic shear waves \citep{Stokes_2009, Ferry1942, Pascal1989, Ortn2020}.

This study considers a glass--silicon--glass microchannel of width \( W \) and depth \( D \), shown in Fig \ref{fig:Fig_5_2} . The rectangular channel is filled with a viscoelastic fluid characterized by a total viscosity \( \mu = \mu_p + \mu_s \), where \( \mu_p \) and \( \mu_s \) denote the polymeric and solvent contributions, respectively, and a relaxation time \( \tau \). Actuation of the piezoelectric transducer generates a standing acoustic wave along the channel width (along \( X \)-axis), corresponding to a half-wave resonance condition, \( W = \lambda_0/2 \), where \( \lambda_0 \) is the wavelength of the standing wave as illustrated in Fig.~\ref{fig:Fig_5_2}(a). This acoustic excitation perturbs the fluid, inducing spatial variations in the pressure \( p \), density \( \rho \), and velocity field \( \boldsymbol{v} \)~\citep{Bruus2012}. To describe the response of the viscoelastic medium under such excitation, the Oldroyd-B constitutive model~\citep{oldroyd1950} is used.

\section{Theoretical formulation}
\label{sec:Theory}

\subsection{Viscoelastic fluid modeling and perturbations}\label{sec:2.1}
To analyze the characteristics of acoustic streaming in a viscoelastic fluid, the formulation begins with the fundamental governing equations, the continuity and momentum equations. The present framework is restricted to dilute polymer solutions for which shear-thinning effects are negligible. Under these conditions, the Oldroyd-B constitutive model \citep{oldroyd1950} is employed, as it provides a minimal yet effective description of fluids exhibiting both viscous and elastic responses. This model has been widely used to represent the rheological behavior of dilute polymer solutions and biological fluids in acoustofluidic and related flow configurations \citep{Doinikov2021a,Doinikov2021b,Sujith2024,Vargas2022}. Accordingly, the governing equations for a viscoelastic fluid are expressed as follows:

%
%
%
\begin{figure}
\centering
\includegraphics[clip,width=0.95\textwidth]{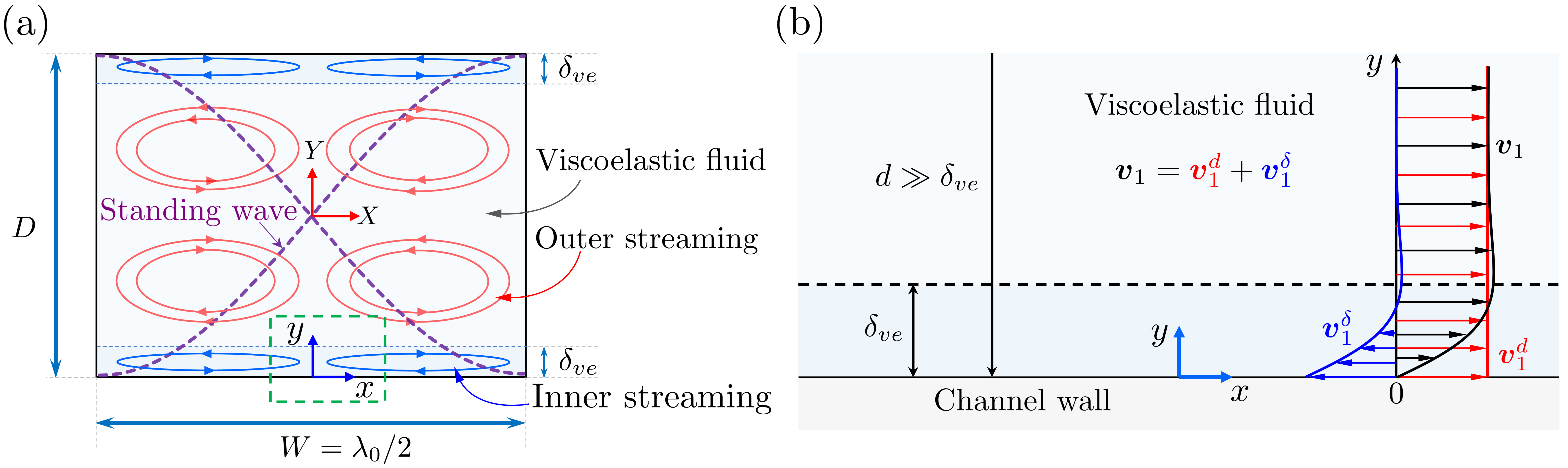}
\caption
{
\justifying{A sketch of the section of viscoelastic fluid and stationary channel wall with acoustic boundary layer formation. The bulk fluid supports a horizontal 1-D standing sinusoidal pressure wave of wavelength $\lambda_0$ along the $x$ axis of channel. The interaction between oscillating viscoelastic fluid and the stationary wall forms a acoustic boundary layer closed to the wall, represented by $\delta_{ve}$, indicate the short range fluid region with an incompressible solenoidal flow. The region away from the $\delta_{ve}$ is long range field with compressible potential flow, represented by $d$. Using Helmholtz decomposition the total first order fluid velocity is consider as $\boldsymbol{v}_1=\boldsymbol{v}_1^d+\boldsymbol{v}_1^\delta$. A schematic variation of $\boldsymbol{v}_1^d$  (red),
$\boldsymbol{v}_1^\delta$ (blue) and
$\boldsymbol{v}_1$  (black) is shown with arrows, which indicate the direction of field.}
}
\label{fig:Fig_5_2}
\end{figure}
\begin{equation}\label{eq:5_1}
\frac{\partial \rho}{\partial t}+\boldsymbol{\nabla} \cdot(\rho \boldsymbol{v})=0
\end{equation}
and
\begin{equation}\label{eq:5_2}
\rho\left(\frac{\partial \boldsymbol{v}}{\partial t}+\boldsymbol{v} \cdot \boldsymbol{\nabla} \boldsymbol{v}\right) =-\boldsymbol{\nabla} p+  \boldsymbol{\nabla} \cdot \mu_s\left[\boldsymbol{\nabla} \boldsymbol{v}+(\boldsymbol{\nabla} \boldsymbol{v})^T-\frac{2}{3}(\boldsymbol{\nabla} \cdot \boldsymbol{v}) \boldsymbol{I}\right]+\boldsymbol{\nabla} \cdot \boldsymbol{\sigma}.
\end{equation}
\noindent The fluid velocity is expressed as $\boldsymbol{v}=v_x \boldsymbol{e}_x+v_y\boldsymbol{e}_y$, where $\boldsymbol{e}_x$ and $\boldsymbol{e}_y$ are unit vectors along the $x$ and $y$ directions, and $\mu_s$ denotes the solvent viscosity. In Eq.~\ref{eq:5_2}, consistent with Refs.~\cite{Doinikov2021a,Doinikov2021b,Vargas2022}, for dilute polymer solutions used in the present work, acoustic energy dissipation is dominated by shear viscosity, particularly in Rayleigh streaming near solid boundaries. Accordingly, the role of bulk viscosity is taken to be negligible. For concentrated polymer solutions, however, frequency-dependent bulk viscosity may become relevant, highlighting the need for improved experimental techniques and offering a potential avenue for future investigation. 

In Eq.(\ref{eq:5_2}), $\boldsymbol{\sigma}$ represents the viscoelastic stress tensor corresponds to the Oldroyd-B model \citep{oldroyd1950}, and is given by
\begin{equation}\label{eq:5_3}
\tau \overset{\nabla}{\boldsymbol{\sigma}}+\boldsymbol{\sigma}=\mu_p\left[\boldsymbol{\nabla} \boldsymbol{v}+(\boldsymbol{\nabla} \boldsymbol{v})^T-\frac{2}{3}(\boldsymbol{\nabla} \cdot \boldsymbol{v}) \boldsymbol{I}\right].
\end{equation}

\noindent Here, $\mu_p$ denotes the polymer viscosity, $\tau$ is the relaxation time, and $\overset{\nabla}{\boldsymbol{\sigma}}$ stands for the upper-convected stress derivative of ${\boldsymbol{\sigma}}$, defined as
\begin{equation}\label{eq:5_4}
\overset{\nabla}{\boldsymbol{\sigma}}=\frac{\partial \boldsymbol{\sigma}}{\partial t}+\boldsymbol{v} \cdot \boldsymbol{\nabla} \boldsymbol{\sigma}-\left(\boldsymbol{\nabla} \boldsymbol{v}\right)^T \cdot \boldsymbol{\sigma}-\boldsymbol{\sigma} \cdot \boldsymbol{\nabla} \boldsymbol{v},
\end{equation}
which ensures the frame invariance in the constitutive description of viscoelastic fluids.

To examine fluid fluctuations driven by the acoustic field, a perturbation approach is applied to the governing equations. The relevant fluid variables, density \( \rho \), pressure \( p \), velocity \( \boldsymbol{v} \), and stress tensor \( \boldsymbol{\sigma} \), are expressed as asymptotic expansions in a small perturbation parameter and retained up to second order \citep{Bruus2012b,Bruus2012,Powell1982,Sujith2024}. Substitution of these expansions into the governing equations, followed by collection of terms of equal order, yields a hierarchy of equations. At first order, only linear terms are retained, while all nonlinear products of first-order quantities are neglected. This first-order governing equations can be written as

\begin{equation}\label{eq:5_5}
\frac{\partial \rho_1}{\partial t}+\rho_0 \boldsymbol{\nabla} \cdot \boldsymbol{v}_1=0,
\end{equation}
\begin{equation}\label{eq:5_6}
\rho_0 \frac{\partial \boldsymbol{v}_1}{\partial t}=-\boldsymbol{\nabla} p_1+  \boldsymbol{\nabla} \cdot \mu_s\left[\boldsymbol{\nabla} \boldsymbol{v}_1+\left(\boldsymbol{\nabla} \boldsymbol{v}_1\right)^T-\frac{2}{3}\left(\boldsymbol{\nabla} \cdot \boldsymbol{v}_1\right) \boldsymbol{I}\right]+\boldsymbol{\nabla} \cdot \boldsymbol{\sigma}_1
\end{equation}
and
\begin{equation}\label{eq:5_7}
\tau \frac{\partial \boldsymbol{\sigma}_1}{\partial \mathrm{t}}+\boldsymbol{\sigma}_1=\mu_p\left[\boldsymbol{\nabla} \boldsymbol{v}_1+\left(\boldsymbol{\nabla} \boldsymbol{v}_1\right)^T-\frac{2}{3}\left(\boldsymbol{\nabla} \cdot \boldsymbol{v}_1\right) \boldsymbol{I}\right].
\end{equation}

\noindent Moreover, the equation of state can be expressed as
\begin{equation}\label{eq:5_8}
p_1-p_0=\left(\rho_1-\rho_0\right) c_0^2.
\end{equation}

\noindent \noindent Here, $p_0$, $\rho_0$ are equilibrium pressure and density, $c_0$ represents the speed of sound in the fluid. The harmonic time dependence for the first-order fields is expressed as \( \rho_1 = \hat{\rho}_1(\boldsymbol{r}) e^{i \omega t} \), \( p_1 = \hat{p}_1(\boldsymbol{r}) e^{i \omega t} \), \( \boldsymbol{v}_1 = \hat{\boldsymbol{v}}_1(\boldsymbol{r}) e^{i \omega t} \), and \( \boldsymbol{\sigma}_1 = \hat{\boldsymbol{\sigma}}_1(\boldsymbol{r}) e^{i \omega t} \)~\citep{Doinikov2021a, Sujith2024}. Here, \( \omega = 2\pi f \) is the angular frequency, where \( f \) denotes the resonant frequency of the standing bulk acoustic wave (S-BAW). The quantities \( \hat{\rho}_1, \ \hat{p}_1, \ \hat{\boldsymbol{v}}_1, \ \hat{\boldsymbol{\sigma}}_1 \) represent the time-independent spatial amplitudes of the first-order fields. Henceforth, the notation ``\ \(\hat{} \ \)'' is used to indicate these time-independent amplitudes. Substituting this harmonic form into Eqs.~(\ref{eq:5_5})--(\ref{eq:5_7}) transforms the first-order equations from the time domain to the frequency domain. Upon simplification, the first-order equations becomes
\begin{equation}\label{eq:5_9}
i \omega \hat{\rho}_1+\rho_0 \boldsymbol{\nabla} \cdot \hat{\boldsymbol{v}}_1=0
\end{equation}
and
\begin{equation}\label{eq:5_10}
\rho_0 \frac{\partial \hat{\boldsymbol{v}}_1}{\partial t}=-\boldsymbol{\nabla} \hat{p}_1 +\boldsymbol{\nabla} \cdot \mu_c\left[\boldsymbol{\nabla} \hat{\boldsymbol{v}}_1+\left(\boldsymbol{\nabla} \hat{\boldsymbol{v}}_1\right)^T-\frac{2}{3}\left(\boldsymbol{\nabla} \cdot \hat{\boldsymbol{v}}_1\right) \boldsymbol{I}\right],
\end{equation}
\noindent respectively, where 
\begin{equation}\label{eq:5_11}
\mu_c=\mu_s+\frac{\mu_p}{1+i \omega \lambda}.
\end{equation}
\noindent Here, $\mu_c$ is the complex viscosity, which describes the effect of viscoelastic parameters on the first-order fluid field. 

The first-order fields are purely oscillatory and therefore do not contribute to time-averaged quantities \citep{Bruus2012}. In contrast, the second-order fields generate a finite time-averaged response. Retaining terms up to second order in the perturbation expansions of Eqs.~(\ref{eq:5_1})--(\ref{eq:5_4}), followed by temporal averaging over one acoustic period, yields the governing equations for the time-averaged second-order fields, expressed as
\begin{equation}\label{eq:5_12}
\rho_0 \boldsymbol{\nabla} \cdot\left\langle\boldsymbol{v}_2\right\rangle=-\boldsymbol{\nabla} \cdot\left\langle\rho_1 \boldsymbol{v}_1\right\rangle
\end{equation}
and
\begin{equation}\label{eq:5_13}
\begin{split}
 &\rho_0\left\langle\boldsymbol{v}_1 \cdot \boldsymbol{\nabla} \boldsymbol{v}_1+\boldsymbol{v}_1 \boldsymbol{\nabla} \cdot \boldsymbol{v}_1\right\rangle=-\boldsymbol{\nabla}\left\langle p_2\right\rangle+\mu \boldsymbol{\nabla}^2\left\langle\boldsymbol{v}_2\right\rangle+\frac{\mu}{3} \boldsymbol{\nabla}\left(\boldsymbol{\nabla} \cdot\left\langle\boldsymbol{v}_2\right\rangle\right)\\&
 \ -\tau \boldsymbol{\nabla} \cdot\left\langle\boldsymbol{v}_1 \cdot \boldsymbol{\nabla} \boldsymbol{\sigma}_1\right\rangle+\tau \boldsymbol{\nabla} \cdot\left\langle\left(\boldsymbol{\nabla} \boldsymbol{v}_1\right)^T \cdot \boldsymbol{\sigma}_1\right\rangle+\tau \boldsymbol{\nabla} \cdot\left\langle\boldsymbol{\sigma}_1 \cdot \boldsymbol{\nabla} \boldsymbol{v}_1\right\rangle.
\end{split}
\end{equation}
\noindent The final three terms on the right-hand side of Eq.~(\ref{eq:5_13}) arise from the viscoelastic constitutive model and represent additional body forces introduced by the fluid's viscoelastic nature.

\subsection{Analytical form of boundary-layer driven streaming}\label{sec:2.2}

The analysis is restricted to the cross-section of the microchannel (see Fig.~\ref{fig:Fig_5_2}), where the streaming flow is assumed to be two-dimensional in the \(X\text{-}Y\) plane and uniform along the channel axis (\(Z\)-direction). The channel width is selected as half of the ultrasonic wavelength, ensuring a resonant configuration that produces a one-dimensional standing acoustic wave along the \(X\)-axis. The first-order acoustic velocity field is expressed as \( \boldsymbol{v}_1 = \hat{\boldsymbol{v}}_1 e^{i\omega t} \), with \( \hat{\boldsymbol{v}}_1 = A \omega \cos(k_X X)\,\boldsymbol{e}_X \), where \(A\) denotes the amplitude of wall displacement.

In the vicinity of the channel walls, the oscillatory velocity attenuates due to viscous dissipation and fluid–solid interaction, consistent with the no-slip boundary condition. This attenuation occurs over a characteristic length scale \( \delta_{ve} \), which defines the thickness of the viscoelastic acoustic boundary layer. The problem involves multiple length scales, including the oscillation amplitude \(A\), channel width \(W\), channel depth \(D\), boundary-layer thickness \( \delta_{ve} \), and acoustic wavelength \( \lambda_0 \). For typical acoustofluidic systems, these scales satisfy the ordering \( A \ll \delta_{ve} \ll \lambda_0 \sim W \sim D \). Accordingly, the Stokes parameter is defined as \( S_p = \delta_{ve}/\lambda_0 \ll 1 \), and the thin-boundary-layer assumption \( \delta_{ve}/D \ll 1 \) is adopted.

A standard method for addressing such problem is through boundary-layer theory, which divides the domain into a bulk region and a thin boundary layer near solid surfaces. The first-order velocity is solved in each region and matched asymptotically. Given that \( \delta_{ve} \ll \lambda_0 \), the boundary-layer flow is often assumed incompressible. However, this assumption is only partially valid: the flow is incompressible in the wall-normal direction due to the short scale \( \delta_{ve} \), but remains compressible along the wall, where variations occur over the longer scale \( \lambda_0 \), similar to the bulk. Full incompressibility is valid only in specific cases, such as flow around particles of size \( a \ll \lambda_0 \), where both \( a \) and \( \delta_{ve} \) determine the spatial dynamics \citep{Bruus2012}. To address this limitation, a refined modeling approach is considered \citep{Bach2018}, where the field is decomposed into components, which explicitly identify compressible and incompressible components throughout the domain, enabling a more accurate and complete representation of the flow. As a starting point of the study, the Helmholtz decomposition of the first  order velocity field is consider as
\begin{equation}\label{eq:5_14}
\boldsymbol{v}_1=\boldsymbol{v}_1^d+\boldsymbol{v}_1^\delta.
\end{equation}
\

\noindent Here, $\boldsymbol{v}_1^d$ represents the compressible, irrotational (potential) flow, satisfying $\boldsymbol{\nabla} \times \boldsymbol{v}_1^d = 0$, while $\boldsymbol{v}_1^\delta$ denotes the incompressible, solenoidal flow, with $\boldsymbol{\nabla} \cdot \boldsymbol{v}_1^\delta = 0$. Moreover, $\boldsymbol{v}_1^d=\hat{\boldsymbol{v}}_1^d e^{i \omega t}$, $\boldsymbol{v}_1^{\delta}=\hat{\boldsymbol{v}}_1^{\delta} e^{i \omega t}$, the time-independent spatial amplitudes are denoted by $\hat{\boldsymbol{v}}_1^d$ and $\hat{\boldsymbol{v}}_1^\delta$, respectively. The superscripts ``$d$'' and ``$\delta$'' refer to the bulk and boundary layer regions, as illustrated in Fig.~\ref{fig:Fig_5_2}(b). Applying Eq. (\ref{eq:5_14}) in Eqs.(\ref{eq:5_9}) and (\ref{eq:5_10}), followed by separating the irrotational and solenoidal and further
simplification using Eq. (\ref{eq:5_8}) gives
\begin{equation}\label{eq:5_15}
i \omega \kappa_0 \hat{p}_1=- \boldsymbol{\nabla} \cdot \hat{\boldsymbol{v}}_1^d,
\end{equation}
\begin{equation}\label{eq:5_16}
i \omega \rho_0 \hat{\boldsymbol{v}}_1^d =-
\gamma \boldsymbol{\nabla} \hat{p}_1, \quad \textrm{where} \ \gamma=(1+\frac{4}{3}i \omega \kappa_0 \mu_c),
\end{equation}
and
\begin{equation}\label{eq:5_17}
i \omega \rho_0 \hat{\boldsymbol{v}}_1^\delta=\mu_c \boldsymbol{\nabla}^2 \hat{\boldsymbol{v}}_1^\delta.
\end{equation}
\

\noindent Here, $\kappa_0=1/\rho_0 c_0^2$, from Eqs. (\ref{eq:5_15}), (\ref{eq:5_16}) and (\ref{eq:5_17}), the Helmholtz equation for $p_1$, $\boldsymbol{v}_1^d$, and $\boldsymbol{v}_1^\delta$ can be expressed as
\begin{equation}\label{eq:5_18}
\boldsymbol{\nabla}^2 \hat{p}_1+k_d^2 \hat{p}_1=0,
\end{equation}
\begin{equation}\label{eq:5_19}
\boldsymbol{\nabla}^2 \hat{\boldsymbol{v}}_1^d+k_d^2 \hat{\boldsymbol{v}}_1^d=0,
\end{equation}
and
\begin{equation}\label{eq:5_20}
\boldsymbol{\nabla}^2 \hat{\boldsymbol{v}}_1^\delta- k_s^2 \hat{\boldsymbol{v}}_1^\delta=0.
\end{equation}

\noindent The wave numbers $k_d$ and $k_s$ are given by 
\begin{equation}\label{eq:5_21}
k_d=\frac{k_x}{\gamma^{1/2}} \quad \textrm{and} \quad k_s=\left( \frac{i \omega \rho_0}{\mu_c}\right)^{1/2}.
\end{equation}
Here, \(k_x = 2\pi/\lambda_0\) denotes the acoustic wavenumber, while \(k_s\) is the complex shear wavenumber associated with the viscoelastic fluid. Drawing an analogy with viscous fluids, the acoustic boundary layer thickness for a viscoelastic fluid is defined as \(\delta_{ve} = 1/\textrm{Im}[k_s]\), and the corresponding shear wavelength as \(\lambda_{ve} = 2\pi/\textrm{Re}[k_s]\). For a Newtonian fluid, the ratio \(\lambda_{ve}/\delta_{ve}\) reduces to \(\lambda_v/\delta = 2\pi\), whereas for a viscoelastic fluid, this ratio generally deviates from \(2\pi\), \( \lambda_{ve}/\delta_{ve} \leq 2 \pi\) \citep{Doinikov2021a, Sujith2024}. Since acoustic streaming is driven by boundary layer dynamics, the modified boundary layer thickness and shear wavelength are used throughout the subsequent analysis.

Initially the inner boundary layer streaming is solved to find out a slip velocity condition for outer streaming, which is outlined in the upcoming sections. In Eq. (\ref{eq:5_20}) the Laplacian operator $\boldsymbol{\nabla}^2=\partial_x^2+\partial_y^2$.
In the boundary layer region, as shown in Fig. \ref{fig:Fig_5_2}(b), from scaling $\partial_x \sim k_x=2 \pi/ \lambda$ and $\partial_y \sim 1/\delta_{ve}$, where the derivatives along the standing wave direction ($x$-axis) is very less as compared to the $y$ direction in the boundary layer, therefore this study approximate  $\boldsymbol{\nabla}^2 \approx \partial_y^2$ \citep{Bach2018}. Accordingly, Eq. (\ref{eq:5_20}) for $x$ component of $\boldsymbol{v}_1^\delta$ reduces to
\begin{equation}\label{eq:5_22}
\partial^2_y \hat{v}_{1,x}^\delta=k_s^2 \hat{v}_{1,x}^\delta.
\end{equation}
 Double integration of Eq. (\ref{eq:5_22}) gives
\begin{equation}\label{eq:5_23}
\hat{v}_{1,x}^\delta=c_1 e^{k_s y}+c_2 e^{-k_s y},
\end{equation}
where $c_1$ and $c_2$ are constants of integration. A long range boundary condition is applied: when $y \rightarrow \infty$,  $\hat{v}_{1,x}^\delta \rightarrow 0$ and  no-slip boundary condition: at $y=0$,  $\boldsymbol{v}_{1}=0$ (where $\hat{\boldsymbol{v}}_{1,x}^\delta=-\hat{\boldsymbol{v}}_{1}^d$ ) on Eq. (\ref{eq:5_23}). Which gives  
\begin{equation}\label{eq:5_24}
\hat{v}_{1,x}^\delta=-\hat{v}_1^d e^{-k_s y}.
\end{equation}
Here, the irrotational far field first order velocity field is considered as $\hat{\boldsymbol{v}}_1^d =v_1^{d0} \cos \left(k_x x\right) \boldsymbol{e}_x$. 
Where $v_1^{d0}$ is the amplitude of first order velocity perturbation, $v_1^{d0}=A \omega$.
Substitution of Eq.  (\ref{eq:5_11}) to $k_s$ and further expansion and simplification of Eq. (\ref{eq:5_24}) yields to
\begin{equation}\label{eq:5_25}
   \hat{v}_{1,x}^\delta=-v_1^{d_{0}} \cos \left(k_x x\right)  e^{-\chi y/ \delta_{f}}  e^{-i \xi y/ \delta_{f}}. 
\end{equation}
Here, $\delta_f=\sqrt{2 \mu_f/\rho_0 \omega}$, subscript``$f$" indicate the base Newtonian fluid,  considered as Deionized water, $\chi$ and $\xi$ can be expressed as
\begin{equation}\label{eq:5_26}
\chi=\sqrt{\sqrt{\Gamma^2+\Xi^2}+\Gamma},\ \quad
\xi=\sqrt{\sqrt{\Gamma^2+\Xi^2}-\Gamma}.
\end{equation}
The parameters \( \Gamma \) and \( \Xi \) encapsulate the influence of the fluid's viscoelasticity and are expressed in terms of dimensionless quantities derived from relevant time scales. To systematically characterize the viscoelastic behavior, four fundamental time scales are considered: the acoustic time \( t_{ac} = 1/\omega \), the polymer relaxation time \( \tau \), the polymer viscous diffusion time \( t_{d,p} = (D/2)^2 / 2\nu_p \), and the solvent viscous diffusion time \( t_{d,s} = (D/2)^2 / 2\nu_s \). Here, \( \omega = 2\pi f \) is the angular frequency of the acoustic field, while \( \nu_p \) and \( \nu_s \) denote the kinematic viscosities of the polymer and solvent, respectively. The interplay between these time scales is captured using two key dimensionless groups: the Deborah number, \( De = \tau / t_{ac} \), which quantifies the relative importance of elastic to acoustic time scales, and the viscous diffusion number, \( Dv = t_{d,s} / t_{d,p} = \nu_p / \nu_s \), which compares the diffusive behavior of polymer and solvent viscosity components. Additionally, the dimensionless solvent viscosity is defined as \( \mu_s^* = \mu_s / \mu_f \), where \( \mu_f \) is the viscosity of the base Newtonian fluid used as a reference. Finally, $\Gamma$ and $\Xi$ can be expressed as
\begin{equation}\label{eq:5_27}
    \Gamma=-\frac{1}{\mu_s^*}\Biggl\{\frac{Dv\ De}{Dv^{2}+2 Dv+(1+De^2)}\Biggr\}
\end{equation}
and
\begin{equation}\label{eq:5_28}
    \Xi=\frac{1}{\mu_s^*}\Biggl\{\frac{Dv+1+De^2 }{Dv^{2}+2 Dv+(1+De^2)}\Biggr\}.
\end{equation}

As mentioned before, from incompressible solenoidal flow inside the boundary layer $\boldsymbol{\nabla} \boldsymbol \cdot {v}_1^\delta=0$. Using this along with the no slip boundary condition at the surface $\hat{v}_{1,y}^\delta=0$, the $y$ component of first order velocity field inside the boundary layer is obtained,
\begin{equation}\label{eq:5_29}
\hat{v}_{1,y}^\delta=-\frac{\chi-i \xi}{\chi^2+\xi^2} v_1^{d_{0}} \delta_{f} k_x \sin \left(k_x x\right)  (1-e^{-\chi y/ \delta_{f}}  e^{-i \xi y/ \delta_{f}}).    
\end{equation}
Considering equations Eqs. (\ref{eq:5_14}), (\ref{eq:5_25}), and (\ref{eq:5_29}); the total first order velocity solutions becomes
\begin{equation}\label{eq:5_30}
   {v}_{1,x}=v_1^{d_{0}} \cos \left(k_x x\right) (1- e^{-\chi y/ \delta_{f}}  e^{-i \xi y/ \delta_{f}}) \ e^{i \omega t}
\end{equation}
and
\begin{equation}\label{eq:5_31}
{v}_{1,y}=-\frac{\chi-i \xi}{\chi^2+\xi^2} v_1^{d_{0}} \delta_{f} k_x \sin \left(k_x x\right)  (1-e^{-\chi y/ \delta_{f}}  e^{-i \xi y/ \delta_{f}}) \ e^{i \omega t}.    
\end{equation}
Here, $\delta_f k_x \ll 1$, therefore, the $y$-component of the first-order velocity, $\hat{v}_{1,y}^\delta$ and $v_{1,y}$ becomes negligible as compared to the $x$-component. 

To analyze acoustic streaming, second-order fluid fields is decomposed into two distinct components corresponding to their characteristic spatial scales: a boundary layer component governed by the length scale \( \delta_{ve} \), and a bulk component associated with \( D \sim \mathcal{O}(\lambda_0) \). The short-range boundary layer field is denoted by \( \boldsymbol{f}_2^\delta \), and the long-range bulk fields by \( \boldsymbol{f}_2^d \) \citep{Bach2018} and $\epsilon=\delta/d \ll 1 $. The total second-order field is then expressed as the sum of these contributions,
\begin{equation}\label{eq:5_32}
\boldsymbol{f}_2=\boldsymbol{f}^d_2+\boldsymbol{f}^\delta_2.
\end{equation}
Substitution of Eq. (\ref{eq:5_32}) for second order pressure, velocity, and stress in Eqs. (\ref{eq:5_12}) and (\ref{eq:5_13})  and separating short range field terms yields to
\begin{equation}\label{eq:5_33}
\rho_0 \boldsymbol{\nabla} \cdot\left\langle\boldsymbol{v}_2^\delta \right\rangle=-\boldsymbol{\nabla} \cdot\left\langle\rho_1 \boldsymbol{v}_1^\delta \right\rangle
\end{equation}
and
\begin{multline}\label{eq:5_34}
\rho_0 \boldsymbol{\nabla} \cdot \left\langle\boldsymbol{v}_1^\delta \boldsymbol{v}_1^\delta+ \boldsymbol{v}_1^\delta \boldsymbol{v}_1^d+\boldsymbol{v}_1^d \boldsymbol{v}_1^\delta\right\rangle=-\boldsymbol{\nabla}\left\langle p_2^\delta \right\rangle+\mu {\nabla}^2\left\langle\boldsymbol{v}_2^\delta \right\rangle +\frac{\mu}{3} \boldsymbol{\nabla}\left(\boldsymbol{\nabla} \cdot\left\langle\boldsymbol{v}_2^\delta \right\rangle\right) \\ -
\tau\boldsymbol{\nabla} \cdot\left\langle\boldsymbol{v}_1^\delta \cdot \boldsymbol{\nabla} \boldsymbol{\sigma}_1\right\rangle  +
\tau \boldsymbol{\nabla} \cdot\left\langle\left(\boldsymbol{\nabla} \boldsymbol{v}_1^\delta\right)^T \cdot \boldsymbol{\sigma}_1\right\rangle+ \tau \boldsymbol{\nabla} \cdot\left\langle\boldsymbol{\sigma}_1 \cdot \boldsymbol{\nabla} \boldsymbol{v}_1^\delta\right. \left.\right\rangle.
\end{multline}

To gain insight into the relative significance of the terms in Eq.~(\ref{eq:5_34}),a scaling analysis is performed. As a first step, the intial focus is on the pressure component \( p_2^\delta \) by taking the divergence of Eq.~(\ref{eq:5_34}). This expression is then simplified using Eqs.~(\ref{eq:5_33}) and (\ref{eq:5_15}), along with the incompressibility condition \( \boldsymbol{\nabla} \cdot \boldsymbol{v}_1^\delta = 0 \). The resulting equation for \( p_2^\delta \) can be written as
\begin{multline}\label{eq:5_35}
    {\nabla}^2 \left\langle p_2^\delta \right\rangle=-\rho_0 \boldsymbol{\nabla} \cdot\left(\boldsymbol{\nabla} \cdot \left\langle\boldsymbol{v}_1^\delta \boldsymbol{v}_1^\delta+ \boldsymbol{v}_1^\delta \boldsymbol{v}_1^d+\boldsymbol{v}_1^d \boldsymbol{v}_1^\delta\right\rangle\right)+\frac{4}{3i \omega}\mu k_d^2 {\nabla}^2 \left\langle \boldsymbol{v}_1^\delta \cdot  \boldsymbol{v}_1^d \right\rangle-\\ 
    \tau \boldsymbol  {\nabla}\cdot \left(\boldsymbol{\nabla} \cdot\left\langle\boldsymbol{v}_1^\delta \cdot \boldsymbol{\nabla} \boldsymbol{\sigma}_1\right\rangle+  \boldsymbol{\nabla} \cdot\left\langle\left(\boldsymbol{\nabla} \boldsymbol{v}_1^\delta\right)^T \cdot \boldsymbol{\sigma}_1\right\rangle +    \boldsymbol{\nabla} \cdot\left\langle\boldsymbol{\sigma}_1 \cdot \boldsymbol{\nabla} \boldsymbol{v}_1^\delta\right\rangle\right)
\end{multline}

In Eq.~(\ref{eq:5_35}), the first, second, and third terms on the right-hand side arise from Reynolds stress (RES), viscous stress (V), and viscoelastic stress (VES), respectively. These contributions are denoted as \( T_{\textrm{RES}} \), \( T_{\textrm{V}} \), and \( T_{\textrm{VES}} \). To estimate their magnitudes, the following scalings are applied: 

\begin{equation}
    \lvert v_{1,x}^{\delta} \rvert \rightarrow v_1^{d_0}, \  \lvert v_{1,y}^{\delta} \rvert \rightarrow \epsilon v_1^{d_0}, \ \lvert v_{1,x}^{d} \rvert \rightarrow v_1^{d_0}, \ \lvert v_{1,y}^{d} \rvert \rightarrow 0, \ \lvert v_{1,x} \rvert \rightarrow v_1^{d_0}, \ \lvert v_{1,y} \rvert \rightarrow \epsilon v_1^{d_0},
\end{equation}
where \( \epsilon = \delta / d \ll 1 \). Based on the dominant contributions, the leading-order scalings of the terms are:
\begin{equation}
   \lvert T_{\textrm{RES}} \rvert \sim \rho_0 d^{-2} (v_1^{d_0})^2, \quad \lvert T_{\textrm{V}} \rvert \sim \frac{\mu}{\omega} (\delta d)^{-2} (v_1^{d_0})^2, \quad \lvert T_{\textrm{VES}} \rvert \sim \tau \mu (\delta d)^{-2} (v_1^{d_0})^2. 
\end{equation}
The scaling of the second-order pressure term yields: $\lvert \nabla^2 \langle p_2^\delta \rangle \rvert \sim \delta^{-2} p_2^\delta$. By equating both sides of Eq.~(\ref{eq:5_35}),  the scaling for the pressure can be obtained as
\begin{equation}
    \lvert p_2^\delta \rvert \sim \epsilon^2 \rho_0 (v_1^{d_0})^2, \quad \frac{\mu}{\omega} d^{-2} (v_1^{d_0})^2, \quad \tau \mu d^{-2} (v_1^{d_0})^2.
\end{equation}
Since \( d^{-2} \) and \( \epsilon \) are small, the pressure gradient \( \boldsymbol{\nabla} p_2^\delta \) becomes negligible and is omitted from further analysis of the boundary layer in Eq.~(\ref{eq:5_34}). Additionally, the term \( (\mu/3) \boldsymbol{\nabla} (\boldsymbol{\nabla} \cdot \langle \boldsymbol{v}_2^\delta \rangle) \) in Eq.~(\ref{eq:5_34}) simplifies to \( (\mu/3\rho_0) \boldsymbol{\nabla} \langle \boldsymbol{v}_1^\delta \cdot \boldsymbol{\nabla} \rho_1 \rangle \), which scales as: $\mu d^{-2} (v_1^{d_0})^2/3c_0$, which is very less than $\lvert \mu \nabla^2 \boldsymbol{v}_2^\delta \rvert$  $\sim \mu \delta^{-2} (v_1^{d_0})^2/c_0$ \citep{Bach2018}. Therefore, further simplification and rearrangement of Eq. (\ref{eq:5_34}) gives
\begin{multline}\label{eq:5_39}
\mu {\nabla}^2 \left\langle \boldsymbol{v}_2^\delta\right\rangle=\rho_0 \boldsymbol{\nabla} \cdot \left\langle\boldsymbol{v}_1^\delta \boldsymbol{v}_1^\delta+ \boldsymbol{v}_1^\delta \boldsymbol{v}_1^d+\boldsymbol{v}_1^d \boldsymbol{v}_1^\delta\right\rangle +
\tau\boldsymbol{\nabla} \cdot\left\langle\boldsymbol{v}_1^\delta \cdot \boldsymbol{\nabla} \boldsymbol{\sigma}_1\right\rangle  \\ -
\tau \boldsymbol{\nabla} \cdot\left\langle\left(\boldsymbol{\nabla} \boldsymbol{v}_1^\delta\right)^T \cdot \boldsymbol{\sigma}_1\right\rangle- \tau \boldsymbol{\nabla} \cdot\left\langle\boldsymbol{\sigma}_1 \cdot \boldsymbol{\nabla} \boldsymbol{v}_1^\delta\right. \left.\right\rangle.
\end{multline}
From Eq. (\ref{eq:5_39}) an analytical solution of $\langle v_{2,x}^\delta \rangle$ can be obtain by integrating it twice and by implying the boundary conditions $y \rightarrow \infty$, $\langle v_{2,x}^\delta \rangle \rightarrow 0$, $ \partial \langle v_{2,x}^\delta \rangle/\partial y \rightarrow 0$. Here $\nabla^2 \approx \partial_y^2$, therefore the component of time averaged second order streaming velocity along the standing wave becomes
\begin{multline}\label{eq:5_40}
   \langle v_{2,x}^\delta \rangle=-\frac{3}{8} \frac{(v_1^{d_0})^2}{c_0} \sin[2k_x x] \frac{1}{3 (1+De^2) \mu_s (Dv+1) \chi^2 (\chi^2+\xi^2)^2} e^{-{2\chi y}/{\delta_f}} \\ \biggl\{ \xi (\chi^2+\xi^2) \ J_1+   2  e^{{\chi y}/{\delta_f}} \chi^2 \ J_2 \ \cos{[\xi y/\delta_{f}]}-2  e^{{\chi y}/{\delta_f}} \chi^2 \ J_3 \ \sin{[\xi y/\delta_{f}]}\biggr\}.
\end{multline}
Here $J_1, J_2$ and $J_3$ can be defined as
\begin{multline}
    J_1=2 \ De \ \mu_s  Dv \ \xi \bigl(\delta_f^2 \ k_x^2+\chi^2-\xi^2\bigr)+  \delta_f^2 \ \rho_0\omega \xi  +\\  De^2 \ \Bigl\{-4 \ \mu_s Dv \ \chi^3+ \delta_f^2 \bigl(4 k_x^2 \mu_s Dv \ \chi+\rho_0\omega \xi \bigr)\Bigr\},
\end{multline}
\begin{multline}
    J_2=De \mu_s Dv\bigl(\chi^4+4 \ \delta_f^2 \ k_x^2 \ \xi^2-\xi^4\bigr)-2 \delta_f^2 \ \rho_0 \omega \bigl(\chi^2-2 \xi^2\bigr) + \\ De^2\biggl\{2 \mu_s Dv\chi \xi (\chi^2+\xi^2)-2 \delta_f^2 \Bigl(2 k_x^2 \mu_s Dv\chi \xi+ \rho_0 \omega \bigl(\chi^2-2 \xi^2\bigr) \Bigr)\biggr\},
\end{multline}
\begin{multline}
    J_3=-2 \ De \mu_s Dv \chi \xi \bigl(2 \delta_f^2 \ k_x^2+\chi^2+\xi^2)-6 \delta_f^2 \rho_0 \omega \chi \xi - \\ De^2\biggl\{ \mu_s Dv(-\chi^4+\xi^4)+2 \delta_f^2 \Bigl(k_x^2 \mu_s Dv(\chi^2+ 3\xi^2)+ 3 \rho_0 \omega \chi \xi \Bigr)\biggr\}.
\end{multline}
In the present analysis, \( \epsilon = \delta / d \ll 1 \) implies that \( \langle v_{2,y}^\delta \rangle  \sim \epsilon \langle v_{2,x}^\delta \rangle \), giving transverse component of the second-order streaming velocity in the boundary layer negligible. Nevertheless, for completeness and to provide a comprehensive theoretical framework, an explicit expression for \( \langle v_{2,y}^\delta \rangle \) is derived. Using Eqs. (\ref{eq:5_33}), 
 (\ref{eq:5_8}), (\ref{eq:5_15}), 
 (\ref{eq:5_25}) and (\ref{eq:5_40}),  the component of second order streaming velocity perpendicular the standing wave becomes 
\begin{multline}\label{eq:5_44}
    \langle v_{2,y}^\delta \rangle=-\frac{3}{8} \frac{(v_1^{d_0})^2}{c_0}  \frac{k_x \delta_f}{3 (1+De^2) \mu_s (Dv+1) \chi^3 (\chi^2+\xi^2)^3} e^{-{2\chi y}/{\delta_f}} \\ \biggl\{  H_1 \ e^{{\chi y}/{\delta_f}} \ \Bigl(\xi \cos{[\xi y/\delta_f]}+ \chi \sin{[\xi y/\delta_f]}\Bigr)+ \\ \cos{[2 k_x x]} \Bigl( H_2+ H_3 \ e^{{\chi y}/{\delta_f}} \cos{[\xi y/\delta_f]} -H_4 \ e^{{\chi y}/{\delta_f}} \sin{[\xi y/\delta_f]} \Bigr) \biggr\}.
\end{multline}
Here $H_1, \ H_2, \ H_3$ and $H_4$ are
\begin{equation}
     H_1=-2 \ (1+De^2) \ \mu_s (Dv+1) \chi^3 (\chi^2+\xi^2)^2,
\end{equation}
\begin{multline}
     H_2=\xi \bigl(\chi^2+\xi^2\bigr)^2 \ \biggl\{2 De \mu_s Dv\xi \bigl(\delta_f^2 k_x^2+\chi^2-\xi^2\bigr)+c_0 \delta_f^2 k_x \xi \rho_0+ \\ De^2 \Bigl(-4 \mu_s Dv\chi^3+\delta_f^2 k_x \bigl(4 k_x \mu_s Dv\chi+c_0 \xi \rho_0\bigr)\Bigr)\biggr\},
\end{multline}
\begin{multline}
    H_3=2 \chi^3 \ \biggl\{- \mu_s (Dv+1) \xi   \bigl(\chi^2+\xi^2\bigr)^2+2 De \mu_s Dv\chi \bigl( \chi^4+8 \delta_f^2 k_x^2 \xi^2+2 \chi^2 \xi^2+\xi^4\bigr)- \\ 4 c_0 \delta_f^2 k_x \chi \bigl( \chi^2-5 \xi^2\bigr) \rho_0 - De^2 \Bigl( (1-Dv)\mu_s \xi \bigl( \chi^2+\xi^2\bigr)+4 \delta_f^2 k_x \bigl(k_x \mu_s Dv\xi (\chi^2-3 \xi^2)+ \\ c_0 \chi (\chi^2-5\xi^2)\rho_0 \bigr) \Bigr) \biggr\}
\end{multline}
\begin{multline}
    H_4=2 \chi^3 \ \biggl\{ \mu_s (Dv+1) \chi   \bigl(\chi^2+\xi^2\bigr)^2-2 De \mu_s Dv\xi \bigl( 4 \delta_f^2 k_x^2 (\chi^2-\xi^2)+ (\chi^2+\xi^2)^2 \bigr)+ \\ 8 c_0 \delta_f^2 k_x \xi \bigl( -2\chi^2+ \xi^2\bigr) \rho_0 + De^2 \Bigl( (3Dv+1)\mu_s \chi \bigl( \chi^2+\xi^2\bigr)-4 \delta_f^2 k_x \bigl(k_x \mu_s Dv\chi (\chi^2+5 \xi^2)- \\ 2c_0 \xi (-2\chi^2+\xi^2)\rho_0 \bigr) \Bigr) \biggr\}.
\end{multline}

Next the study focuses on $\langle v_{2,x}^\delta \rangle$ to determine the boundary conditions for outer streaming. At wall $y=0$, from Eq.(\ref{eq:5_40}), the second order streaming velocity along the standing wave can be obtained as
\begin{equation}\label{eq:5_49}
    \langle v_{2,x}^\delta \rangle=-\frac{3}{8} \frac{(v_1^{d_0})^2}{c_0} \sin[2k_xx] C_s,
\end{equation}
where $C_s$ is the coefficient of correction   in acoustic streaming due to viscoelasticity of the fluid, which is named as streaming coefficient,
\begin{equation}\label{eq:5_50}
 C_s=\frac{2 \rho_0 c_0^2 I_1+2 De  Dv \mu_s^* I_2+ 2 De^2 I_3}{3(1+De^2)\chi^2 (\chi^2+\xi^2)^2 \rho_0 c_0^2  \mu_s^*(1+Dv)}. 
\end{equation}
\noindent Here $I_1,I_2$ and $I_3$ are functions of $Dv$ and $De$, can be defined as
\begin{equation}\label{eq:5_51}
I_1=-4 \chi^4+9 \chi^2 \xi^2+\xi^4,
\end{equation}
\begin{equation}\label{eq:5_52}
I_2=2 \mu_f \omega \xi^2 (5 \chi^2+\xi^2)+\rho_0 c_0^2(\chi^2-\xi^2)(\chi^2+\xi^2)^2,
\end{equation}
\begin{equation}\label{eq:5_53}
I_3=4 Dv \mu_s \omega \chi \xi (-\chi^2+\xi^2)+\rho_0 c_0^2 (-4 \chi^4+ 9 \chi^2 \xi^2+ \xi^4).
\end{equation}
The boundary layer streaming velocity $\langle v_{2,x}^\delta \rangle$ provides a slip boundary condition for the bulk acoustic streaming field $\langle v_{2,x}^d \rangle$, where $\langle v_{2,x}\rangle=\langle v_{2,x}^\delta\rangle_s+\langle v_{2,x}^d\rangle=0$. Therefore long ranged field at $y=0$ becomes 
\begin{equation}\label{eq:5_54}
    \langle v_{2,x}^d \rangle=-\langle v_{2,x}^\delta \rangle=\frac{3}{8} \frac{(v_1^{d_0})^2}{c_0} \sin[2k_xx] C_s
\end{equation}
Here $C_s$ is obtained from Eqs. (\ref{eq:5_50})-(\ref{eq:5_53}). For a Newtonian viscous fluid, $De=0, \ Dv=0$ where the streaming coefficieint reduces to $1$. Therefore Eq. (\ref{eq:5_49}) reduces to 
\begin{equation}\label{eq:5_55}
    \langle v_{2,x}^\delta \rangle=-\frac{3}{8} \frac{(v_1^{d_0})^2}{c_0} \sin[2k_xx], 
\end{equation}
which is the classical Rayleigh streaming for Newtonian fluids with streaming velocity amplitude $v_{\textrm{str}}=3 (v_1^{d_0})^2/8 c_0 $.

\subsection{Analytical form of outer boundary-layer streaming}\label{sec:2.3}

The long range continuity and momentum equations can be represented as 
\begin{equation}\label{eq:5_56}
    \boldsymbol{\nabla} \cdot \left[ \rho_0\langle  \boldsymbol{v}_2^d \rangle+ \langle \rho_1 \boldsymbol{v}_1^d  \rangle\right]=0
\end{equation}
and
\begin{multline}\label{eq:5_57}
\rho_0 \boldsymbol{\nabla} \cdot \left\langle\boldsymbol{v}_1^d \boldsymbol{v}_1^d\right\rangle=-\boldsymbol{\nabla}\left\langle p_2^d \right\rangle+\mu {\nabla}^2\left\langle\boldsymbol{v}_2^d \right\rangle +\frac{\mu}{3} \boldsymbol{\nabla}\left(\boldsymbol{\nabla}\cdot\left\langle\boldsymbol{v}_2^d\right\rangle\right) -
\tau\boldsymbol{\nabla} \cdot\left\langle\boldsymbol{v}_1^d \cdot \boldsymbol{\nabla} \boldsymbol{\sigma}_1\right\rangle  \\ +
\tau \boldsymbol{\nabla} \cdot\left\langle\left(\boldsymbol{\nabla} \boldsymbol{v}_1^d\right)^T \cdot \boldsymbol{\sigma}_1\right\rangle+ \tau \boldsymbol{\nabla} \cdot\left\langle\boldsymbol{\sigma}_1 \cdot \boldsymbol{\nabla} \boldsymbol{v}_1^d\right. \left.\right\rangle.
\end{multline}
Equation (\ref{eq:5_56}) can be rewritten as
\begin{equation}\label{eq:5_58}
    \boldsymbol{\nabla} \cdot \langle \boldsymbol{v}_2^d \rangle=- \frac{\boldsymbol{\nabla} \cdot  \langle \rho_1 \boldsymbol{v}_1^d  \rangle}{\rho_0}=- \frac{  \langle \rho_1 \boldsymbol{\nabla} \cdot \boldsymbol{v}_1^d + \boldsymbol{v}_1^d  \cdot \boldsymbol{\nabla}\rho_1 \rangle}{\rho_0}
\end{equation}
Using $p_1=\rho_1 c_0^2$ and Eq. (\ref{eq:5_15}), $\boldsymbol{\nabla} \cdot \langle \boldsymbol{v}_2^d \rangle$ is scaled to $(k/c_0) ( v_1^{d_0})^2 $. Here, $k \ll 1$ and $c_0 \gg 1$, therefore Eq. (\ref{eq:5_58}) reduces to \citep{Lee1990,Bach2018}
\begin{equation}\label{eq:5_59}
    \boldsymbol{\nabla} \cdot\boldsymbol \langle {v}_2^d\rangle \approx 0
\end{equation}
Using Eq. (\ref{eq:5_59}), the Eq. (\ref{eq:5_57}) can be rewritten as
\begin{multline}\label{eq:5_60}
\mu {\nabla}^2\left\langle\boldsymbol{v}_2^d \right\rangle =\rho_0 \boldsymbol{\nabla} \cdot \left\langle\boldsymbol{v}_1^d \boldsymbol{v}_1^d\right\rangle+\boldsymbol{\nabla}\left\langle p_2^d \right\rangle +
\tau\boldsymbol{\nabla} \cdot\left\langle\boldsymbol{v}_1^d \cdot \boldsymbol{\nabla} \boldsymbol{\sigma}_1\right\rangle  \\ -
\tau \boldsymbol{\nabla} \cdot\left\langle\left(\boldsymbol{\nabla} \boldsymbol{v}_1^d\right)^T \cdot \boldsymbol{\sigma}_1\right\rangle-\tau \boldsymbol{\nabla} \cdot\left\langle\boldsymbol{\sigma}_1 \cdot \boldsymbol{\nabla} \boldsymbol{v}_1^d\right. \left.\right\rangle.
\end{multline}
As mentioned earlier the first order velocity field is $\boldsymbol{v}_1^d=v_1^{d0} \cos \left(k_x x\right) \boldsymbol{e}_x$. Analysis of curl of right hand side terms of Eq. (\ref{eq:5_60}) gives: $\boldsymbol{\nabla} \times \left[\tau\boldsymbol{\nabla} \cdot\left\langle\boldsymbol{v}_1^d \cdot \boldsymbol{\nabla} \boldsymbol{\sigma}_1\right\rangle -
\tau \boldsymbol{\nabla} \cdot\left\langle\left(\boldsymbol{\nabla} \boldsymbol{v}_1^d\right)^T \cdot \boldsymbol{\sigma}_1\right\rangle- \right. \\ \left. \tau \boldsymbol{\nabla} \cdot\left\langle\boldsymbol{\sigma}_1 \cdot \boldsymbol{\nabla} \boldsymbol{v}_1^d\right. \left.\right\rangle \right] \rightarrow 0$, $ \boldsymbol{\nabla} \times \left[ \rho_0 \boldsymbol{\nabla} \cdot \left\langle\boldsymbol{v}_1^d \boldsymbol{v}_1^d\right\rangle \right]\rightarrow 0$, $\boldsymbol{\nabla} \times \boldsymbol{\nabla} \langle p_2^d\rangle \rightarrow 0$. Therefore the entire right side terms can be represented as the gradient of a scalar function $P$ \citep{Lee1990}, where $P$  is the effective viscoelastic hydrostatic pressure, which consider the effect of momentum flux, second order pressure and viscoelastic stress. Therefore Eq. (\ref{eq:5_60}) can be written as
\begin{equation}\label{eq:5_61}
    \mu \nabla^2 \langle \boldsymbol{v}_2^d\rangle= \boldsymbol{\nabla} P.
\end{equation}
To derive the equations of motion for steady outer streaming, the study considers Eqs. (\ref{eq:5_61}) and (\ref{eq:5_59}) in conjunction with the relevant boundary conditions. On the channel walls, perpendicular to the standing wave, viscoelastic effects give rise to a tangential velocity due to the inner streaming. This leads to the emergence of the bulk second-order streaming slip velocity, where  $\langle \boldsymbol{v}_{2}^d \rangle_s= \langle \boldsymbol{v}_{2,x}^d \rangle $ and  $\langle \boldsymbol{v}_{2}^{\delta} \rangle_s= \langle \boldsymbol{v}_{2,x}^{\delta} \rangle $, can be represented as
\begin{equation}\label{eq:5_62}
   \langle \boldsymbol{v}_{2}^d \rangle_s=  -\langle \boldsymbol{v}_{2}^{\delta} \rangle_s=\frac{3}{8} \frac{(v_1^{d_0})^2}{c_0} \sin[2k_xx] C_s \boldsymbol{e}_x.
\end{equation}
Here, $\langle \boldsymbol{v}_{2}^d \rangle_s$ indicate the modified slip velocity for a viscoelastic fluid for outer streaming. However, on the walls where the the acoustic velocity does not have a tangential component, the no slip boundary condition becomes 
\begin{equation}\label{eq:5_63}
   \langle \boldsymbol{v}_{2}^d \rangle_s=0.
\end{equation}
The bulk streaming velocity is represented as \citep{Lee1990}
\begin{equation}\label{eq:5_64}
\langle \boldsymbol{v}_2^d \rangle=\boldsymbol{\nabla} \times \left( \psi \boldsymbol{e}_z\right),
\end{equation} 
where $\psi$ is the stream function, $\langle \boldsymbol{v}_2^d \rangle=\langle v_{2,x}^d \rangle \boldsymbol{e}_x+\langle v_{2,y}^d\rangle \boldsymbol{e}_y$. Therefore $\langle v_{2,x}^d \rangle= \partial \psi/ \partial y$ and $\langle v_{2,y}^d \rangle= -\partial \psi/ \partial x$. Substituting Eq. (\ref{eq:5_64}) in Eq. (\ref{eq:5_61}) and calculating the curl of the resulting equation gives
\begin{equation}\label{eq:5_65}
{\nabla}^4 \boldsymbol{\psi}=0.
\end{equation} 
As previously noted, $\boldsymbol{\nabla} \times \boldsymbol{\nabla} P = 0$. To obtain the steady streaming patterns, the Eq.~(\ref{eq:5_65}) is solved subject to the appropriate boundary conditions by following \cite{Lee1990}. From Eqs.~(\ref{eq:5_62}) and (\ref{eq:5_63}), $\partial \psi / \partial y = \langle v_{2}^d \rangle_s$ and $\psi = 0$ on the channel walls. For the present case of outer streaming, the coordinate origin is at the center of the channel cross section, with $X$ and $Y$ denoting the directions, as illustrated in Fig.~\ref{fig:Fig_5_2}. Accordingly, at $X = \pm D/2$, the boundary condition becomes $\partial \psi / \partial Y = \langle v_{2}^d \rangle_s$. Solving Eq.~(\ref{eq:5_65}) with these conditions yields
\begin{equation}\label{eq:5_66}
    \psi=\frac{3C_s}{8} \frac{(v_1^{d_0})^2}{c_0}  \sin[2k_xX] \ \Upsilon (k_x,D) \Bigl\{p \sinh[{2k_x Y}]- Y q\cosh[{2k_x Y}] \Bigr\},
\end{equation}
where $\Upsilon (k_x,D)$, $p$ and $q$ are
\begin{equation}\label{eq:5_67}
   \Upsilon (k_x,D)=\frac{1}{ \left( k_x D- \sinh [{k_x D}] \cosh[{k_x D}]\right)}, 
\end{equation}
\begin{equation}\label{eq:5_68}
  p= (D/2) \cosh[{k_x D}], \quad \textrm{and} \ \quad  q= \sinh[{k_x D}].
\end{equation}
Therefore, the components of the streaming velocity can be determined directly from the stream function as: $\langle v_{2,X}^d \rangle= \partial \psi/ \partial Y$ and $\langle v_{2,Y}^d \rangle= -\partial \psi/ \partial X$. Finally, the outer streaming velocity becomes
\begin{equation}\label{eq:2.0071}
    \langle v_{2,X}^d\rangle=\frac{3C_s}{8} \frac{(v_1^{d_0})^2}{c_0}\sin[2 k_x X] \ \Upsilon (k_x,D) \Bigl\{(2 k_x p - q) \cosh[2 k_x Y] - 2 k_x q Y \sinh[2 k_x Y] \Bigr\}
\end{equation}
\begin{equation}\label{eq:2.0072}
    \langle v_{2,Y}^d\rangle=-\frac{3C_s}{8}\frac{(v_1^{d_0})^2}{c_0} \cos[2 k_x X] \ \Upsilon (k_x,D)  \Bigl\{2 k_x p \sinh[{2k_x Y}]- 2 k_x q Y \cosh[{2k_x Y}] \Bigr\}
\end{equation}

The expressions given in Eqs.~(\ref{eq:2.0071}) and (\ref{eq:2.0072}) are derived for a parallel-plate channel and therefore neglect the effects of lateral sidewalls, where the velocity field must satisfy the no-slip condition at \( X = \pm W/2 \). This omission can lead to the presence of a non-zero vertical velocity component \( \langle v_{2,Y}^d \rangle \) near the sidewalls, which must be canceled to maintain physical consistency. To resolve this, an iterative correction procedure outlined in Ref.~\cite{Muller2013} is followed, wherein the problem is corrected by rotating the coordinate system by \(90^\circ\). 
Introducing non-dimensional variables \( \tilde{X} = X/(W/2) \), \( \tilde{Y} = Y/(D/2) \), and the aspect ratio \( \beta = D/W \), the velocity boundary conditions for the rectangular channel become: (i) \( \langle v_{2,X}^d \rangle = \langle v_2^d \rangle_s \) at \( \tilde{Y} = \pm 1 \), (ii) \( \langle v_{2,Y}^d \rangle = 0 \) at \( \tilde{Y} = \pm 1 \), (iii) \( \langle v_{2,X}^d \rangle = 0 \) at \( \tilde{X} = \pm 1 \), and (iv) \( \langle v_{2,Y}^d \rangle = 0 \) at \( \tilde{X} = \pm 1 \). Perpendicular-to-the-wall velocity conditions (ii) and (iii) enforce the vanishing of the velocity component normal to the walls, while parallel-to-the-wall velocity conditions (i) and (iv) specify the tangential components. To simultaneously satisfy all these constraints, an iterative Fourier expansion of the velocity field is employed, expressed as an infinite series, the full formulation can be found in the \cite{Muller2013}. Finally, the outer streaming components in rectangular channel can be represented as
\begin{equation}\label{eq:5_71}
\langle {v}_{2,X}^d(\tilde{X},\tilde{Y})\rangle= C_s \ v_{\textrm{str}} \ \sum_{n=1}^{\infty}\Bigl\{C_{1n}\sin[n\pi\tilde{X}] A^{\parallel}(n\beta,\tilde{Y})+ \\ C_{2n} A^\perp(n\beta^{-1},\tilde{X}) \cos[n\pi\tilde{Y}]\Bigr\},
\end{equation}
\begin{equation}\label{eq:5_72}
\langle {v}_{2,Y}^d (\tilde{X},\tilde{Y})\rangle=C_s \ v_{\textrm{str}} \ \sum_{n=1}^{\infty}\Bigl\{C_{1n}\cos[n\pi\tilde{X}] A^\perp(n\beta,\tilde{Y})+ \\ C_{2n} A^\parallel(n\beta^{-1},\tilde{X}) \sin[n \pi\tilde{Y}]\Bigr\}.
\end{equation}
Here, $v_\textrm{str}=({3}/{8}) ({(v_1^{d_0})^2}/{c_0})$ and the dimensionless outer streaming components $(\langle \tilde{\boldsymbol{v}}_2^d \rangle=\langle \boldsymbol{v}_2^d\rangle/v_\text{str})$ can be represented as

\begin{multline}\label{eq:5_73}
\langle \tilde{v}_{2,X}^d(\tilde{X},\tilde{Y})\rangle= C_s  \ \sum_{n=1}^{\infty}\Bigl\{C_{1n}\sin[n\pi\tilde{X}] A^{\parallel}(n\beta,\tilde{Y})+ \\ C_{2n} A^\perp(n\beta^{-1},\tilde{X}) \cos[n\pi\tilde{Y}]\Bigr\}= C_S \ \langle \tilde{v}_{2,X}^d(\tilde{X},\tilde{Y})\rangle_N ,
\end{multline}
\begin{multline}\label{eq:5_74}
\langle \tilde{v}_{2,Y}^d (\tilde{X},\tilde{Y})\rangle=C_s  \ \sum_{n=1}^{\infty}\Bigl\{C_{1n}\cos[n\pi\tilde{X}] A^\perp(n\beta,\tilde{Y})+ \\ C_{2n} A^\parallel(n\beta^{-1},\tilde{X}) \sin[n \pi\tilde{Y}]\Bigr\}=C_s \ \langle \tilde{v}_{2,Y}^d(\tilde{X},\tilde{Y})\rangle_N.
\end{multline}
In the above equations, $\langle \tilde{v}_{2}^d \rangle_N$ refers to the time-averaged second order classical Rayleigh streaming velocity for a Newtonian fluid in a rectangular channel \citep{Muller2013}. The coefficients $A^\parallel(n\beta,\tilde{Y})$,  and $A^\perp(n\beta,\tilde{Y})$ are obtained as \cite{Muller2013},
\begin{multline}\label{eq:5_75}
A^\parallel(n\beta,\tilde{Y})=\Lambda (n \beta) \bigl\{ n \pi \beta \cosh[n \pi \beta] \cosh[n \pi \beta \tilde{Y}]-\sinh[n \pi \beta] \cosh[n \pi \beta \tilde{Y}] \\ -n \pi \beta \tilde{Y} \sinh[ n \pi \beta] \sinh[n \pi \beta \tilde{y}]\bigr\},
\end{multline}
\begin{multline}\label{eq:5_76}
    A^\perp(n\beta,\tilde{Y})= \Lambda (n \beta) \bigl\{ n \pi \beta \cosh[n \pi \beta] \sinh[n \pi \beta \tilde{Y}]-\sinh[n \pi \beta] \cosh[n \pi \beta \tilde{Y}] \\ -n \pi \beta \tilde{Y} \sinh[ n \pi \beta] \cosh[n \pi \beta \tilde{y}]\bigr\},
\end{multline}
where $\Lambda (n \beta) $ becomes
\begin{equation}\label{eq:5_77}
    \Lambda(n \beta)=\frac{1}{n \pi\beta-\sinh[n\pi\beta]\cosh[n\pi\beta]}.
\end{equation}
Moreover, $A^\parallel(n\beta^{-1},\tilde{X})$, $A^\perp(n\beta^{-1},\tilde{X})$, and $\Lambda(n \beta^{-1})$ are obtained from Eqs. (\ref{eq:5_75}), (\ref{eq:5_76}), and (\ref{eq:5_77}) by replacing $n \beta$ with $n \beta^{-1}$ and $\tilde{Y}$ with $\tilde{X}$. Moreover, $C_{1n}$ and $C_{2n}$ are unknown coefficients and determined using parallel-to-the-wall conditions, followed by \cite{Muller2013},   

\begin{equation}
   \boldsymbol{C}_1 =[\mathbf{I}-\mathbf{A}^\perp(\beta^{-1})\mathbf{A}^\perp(\beta)]^{-1}\cdotp\mathbf{e}_1,
\end{equation}

\begin{equation}
   \boldsymbol{C}_2=-\mathbf{A}^\perp(\beta)\cdot\boldsymbol{C}_1.
\end{equation}

\begin{equation}
    \mathbf{A}_{j,n}^\perp(\beta^{-1})=(-1)^n\int_{-1}^1d\tilde{X} A^\perp(n\beta^{-1},\tilde{X}) \sin(j\pi\tilde{X}).
\end{equation}

Using Eqs.~(\ref{eq:5_40}), (\ref{eq:5_44}), (\ref{eq:5_71}), and (\ref{eq:5_72}), the total second-order streaming velocity is given by \( \langle \boldsymbol{v}_2 \rangle = \langle \boldsymbol{v}_2^{\delta} \rangle + \langle \boldsymbol{v}_2^{d} \rangle \), where \( \langle \boldsymbol{v}_2^{\delta} \rangle \) represents the contribution from the inner boundary layers and \( \langle \boldsymbol{v}_2^{d} \rangle \) corresponds to the bulk (outer) streaming. Since the boundary layer thickness is much smaller than the channel depth \( d \) (i.e., \( \delta \ll d \)), the inner streaming is confined to a narrow region near the top and bottom walls. As a result, the dominant observable flow in the bulk of the fluid is governed by the outer streaming. Therefore, in the bulk region, the total streaming velocity can be approximated as \( \langle \boldsymbol{v}_2 \rangle \approx \langle \boldsymbol{v}_2^{d} \rangle \). The $v_\textrm{str}$ is used to non-dimensionlize the streaming velocity within the channel.

\subsection{Contributions of Reynolds and viscoelastic stresses to acoustic Streaming}\label{sec:2_04}

To quantify the individual contributions of the stresses, the time-averaged second-order streaming slip velocity is expressed as  $\langle \tilde{v}_2^{\delta} \rangle_s = S_{\text{RES},v} + S_{\text{RES},ve} + S_{\text{VES}}$. Here $\langle \tilde{v}_2^{\delta} \rangle_s = \langle {v}_2^{\delta} \rangle_s/ v_\textrm{str}$, from Eqs. (\ref{eq:5_62}) and (\ref{eq:5_50}), the individual contributions of viscous RES (\( S_{\text{RES},v} \)), viscoelastic RES (\( S_{\text{RES},ve} \)), and VES (\( S_{\text{VES}} \)) can be expressed as

\begin{equation}
S_{\text{RES},v} = 
-\frac{ 1}{3} e^{{-2 \chi_v y}/{\delta_f}}
\left(1+
6 e^{{ \chi_v y}/{\delta_f}}
\sin\left[ { \xi_v y}/{\delta_f}\right]
+ 2 e^{{ \chi_v y}/{\delta_f}} 
\cos\left[ { \xi_v y}/{\delta_f}\right]
\right) \sin[2k_x x],
\end{equation}

\begin{multline}
S_{\text{RES},ve} = 
\frac{ 1}{6} 
\Biggl\{\frac{16(\chi^2-2 \xi^2) e^{-\chi y/\delta_f} \cos[\xi y/\delta_f]}{(\chi^2+\xi^2)^2 \mu_s^* (1+Dv)}+
4 e^{{ -\chi_v y}/{\delta_f}}
\cos\left[ { \xi_v y}/{\delta_f}\right]  \\  - \frac{4 \xi e^{{ -2\chi y}/{\delta_f}}}{\chi(\chi^2+\xi^2)^2 \mu_s^* (1+Dv)} \left( \chi^2 \xi+\xi^3+12 \chi^3 e^{\chi y/\delta_f} \sin[\xi y/\delta_f]\right) + \\ 2 e^{{ -2 \chi_v y}/{\delta_f}} \left( 1+6 e^{{  \chi_v y}/{\delta_f}}  \sin[{ \xi_v y}/{\delta_f}]\right) \Biggr\} \sin[2 k_x x],
\end{multline}

\begin{multline}
    S_{\text{VES}}=\frac{2 De Dv}{3 \chi^2 (\chi^2+\xi^2)^2 (1+De^2) (1+Dv)} e^{-2\chi y/\delta_f} \\ \Bigl\{H_1-H_2 e^{\chi y/\delta_f} \cos[\xi y/\delta_f]+H_3 e^{\chi y/\delta_f} \sin[\xi y/\delta_f]\Bigr\} \sin[2 k_x x],
\end{multline}

\begin{equation}
    H_1= \xi \left(\chi^2+\xi^2\right) \left(-\chi^2 \xi+ \xi^3+2 \chi^3 De-2 \chi De \ \delta_f^2 \ k_x^2-\xi \delta_f^2 \ k_x^2\right),
\end{equation}

\begin{equation}
    H_2=\chi^2(\chi^4+2 \chi^3 \xi De+2 \chi \xi De \left(\xi^2-2 \delta_f^2 k_x^2\right)-\xi^4+4 \xi^2 \delta_f^2 k_x^2),
\end{equation}

\begin{equation}
  H_3= \chi^2 (\chi^4 De-2 \chi^3 \xi-2 \chi^2 De \delta_f^2 k_x^2-2 \chi \left(\xi^3+2 \xi \delta_f^2 k_x^2\right)-\xi^2 De \left(\xi^2+6 \delta_f^2 k_x^2\right)).
\end{equation}
Here, $\chi_v$ and $\xi_v$ are functions of viscosity, given by $\chi_v = \xi_v = \left[ \mu_s^*(Dv + 1) \right]^{-0.5}$. In contrast, $\chi$ and $\xi$ depend on both viscosity and elasticity, as defined in Eqs.~(\ref{eq:5_26}), (\ref{eq:5_27}), and (\ref{eq:5_28}).

\section{Experiments}\label{sec:5_3}

\begin{figure}
\centering
\includegraphics[clip,width=0.65\textwidth]{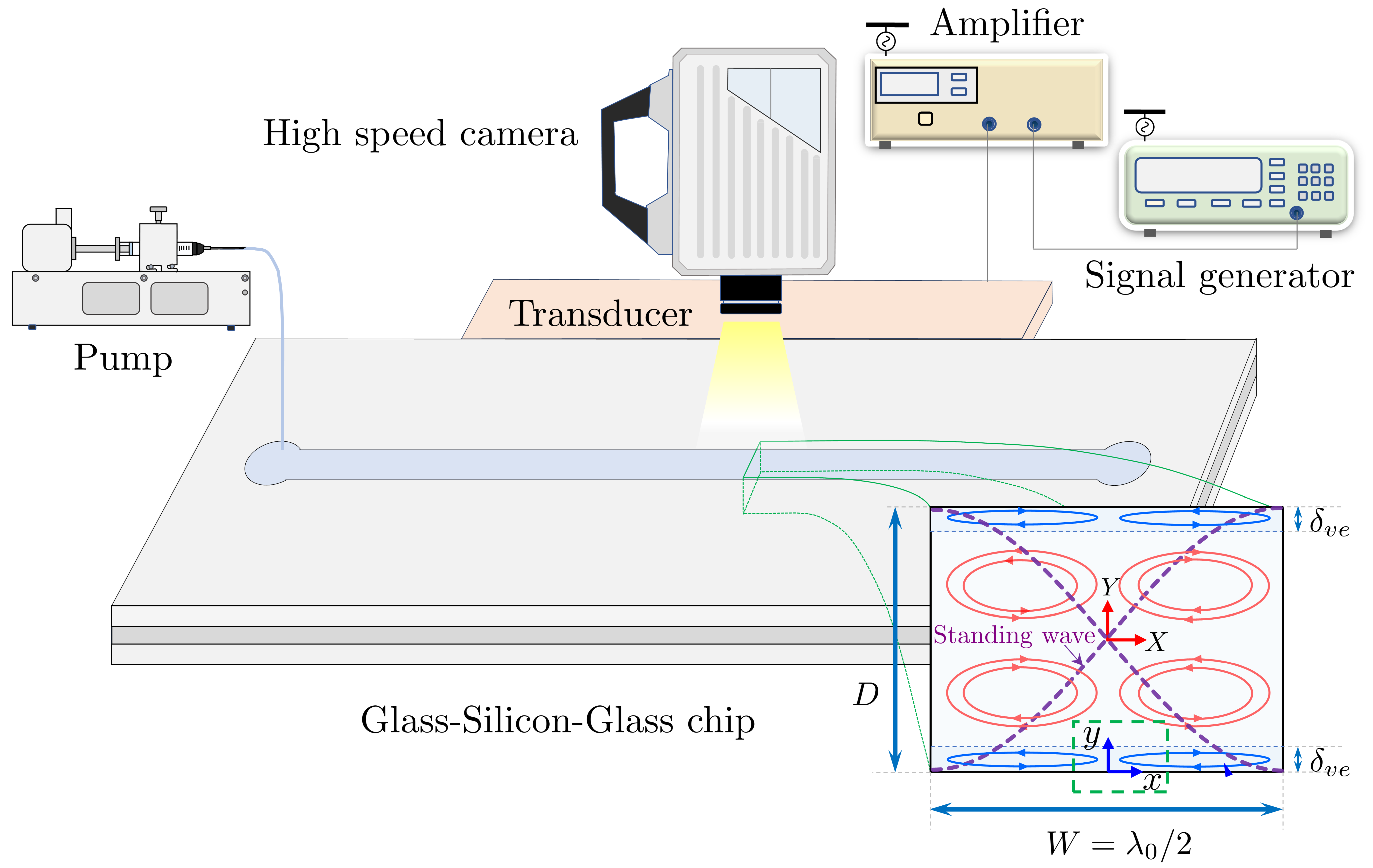}
\caption
{Schematic diagram showing the experimental set up.}
\label{fig:Fig_5_3}
\end{figure}

The experiments are conducted using the same glass–silicon–glass microfluidic chip previously described in \cite{Sujith2024}, also shown in Fig.~\ref{fig:Fig_5_3}. The chip is fabricated via standard photolithography followed by deep reactive ion etching (DRIE). Fabrication begins with a 3-inch, 300~\textmu m thick, $<$100$>$ silicon wafer, onto which the microchannel pattern is transferred using a positive photoresist (MICROPOSIT S1813) and ultraviolet (UV) exposure. The patterned wafer is then through-etched to form a channel with dimensions of 20~mm in length, 400~$\mu$m in width, and 300~$\mu$m in depth. To enclose the etched channel, borosilicate glass plates, each 500~$\mu$m thick, are bonded to both the top and bottom surfaces of the wafer through anodic bonding at 450$^\circ$C under an applied voltage of 1000~V. Inlet and outlet ports are created by micro-drilling holes through the top glass layer to facilitate fluidic access.

The microchannel is filled with either Newtonian or viscoelastic working fluids containing suspended microparticles by a high-precision syringe pump (neMESYS, Cetoni GmbH, Germany), as shown in Fig.~\ref{fig:Fig_5_3}. For acoustic actuation, a piezoelectric transducer (2.0~MHz, Sparkler Piezoceramics) is affixed to the bottom glass layer using epoxy adhesive. It is driven by a sinusoidal signal generated by a function generator (SMC100A, Rohde \& Schwarz, Germany), which is subsequently amplified using a power amplifier (75A250A, Amplifier Research, USA) in the range of 10–1000~mW. The acoustic streaming is visualized using an inverted microscope (IX73, Olympus) with a high-speed CCD camera (Phantom), recording tracer particle motion at 100 frames per second across different focal planes. To determine the resonance frequency of the system, a suspension of 15~$\mu$m polystyrene spheres is introduced into the channel. The transducer is then actuated over a frequency sweep from 1.85 to 2.1~MHz. The resonance condition is identified based on the formation of particle accumulation patterns and the time of migration, with the optimal operating frequency determined to be 1.93~MHz.

\subsection{Types of fluids and characterization}

\begin{figure}
\centering
\includegraphics[clip,width=0.7\textwidth]{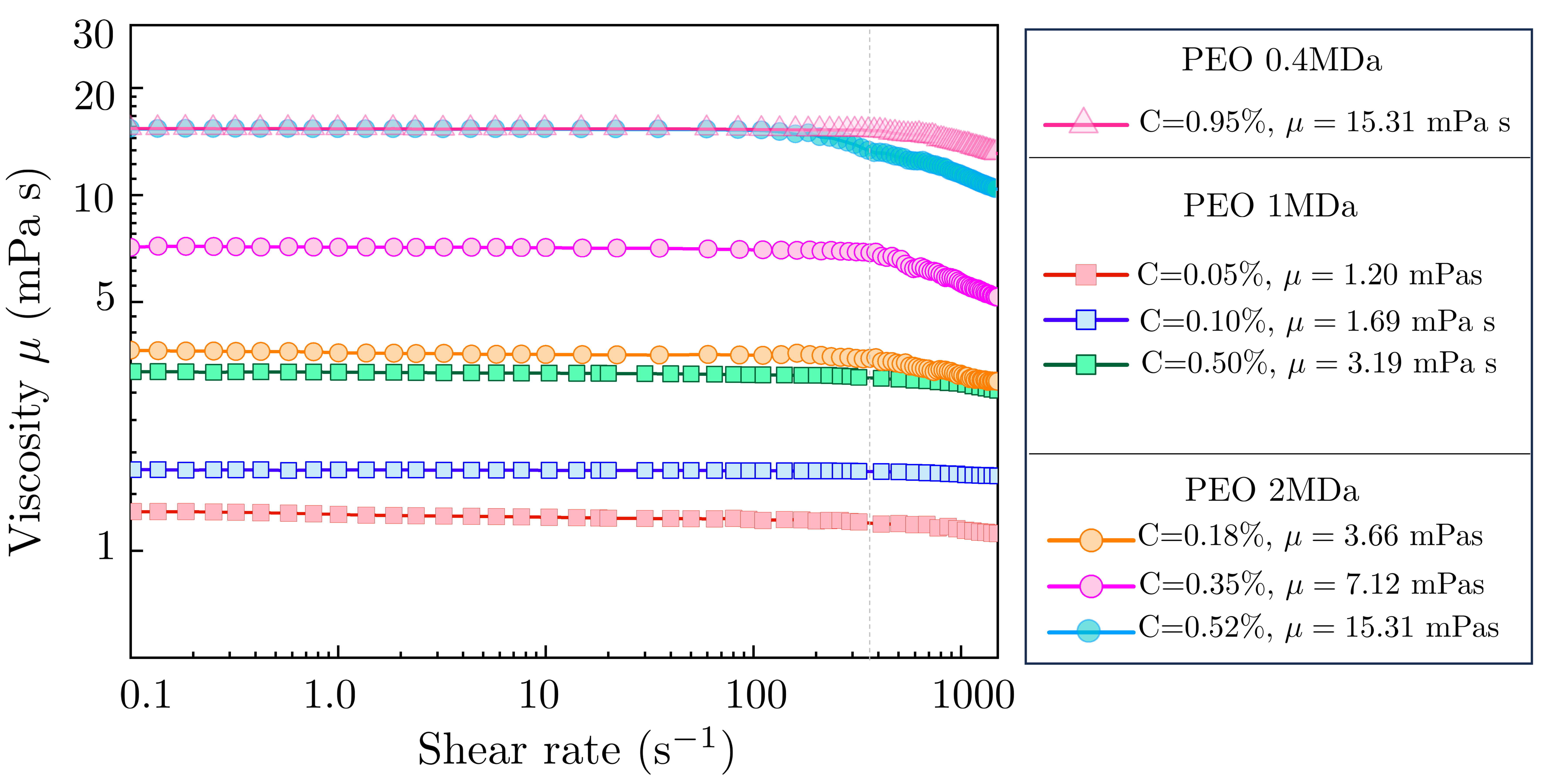}
\caption
{ \justifying Variation of viscosity of aqueous PEO solutions at different concentrations with shear rate, showing a plateau at low shear rates and negligible shear-thinning in the range relevant to acoustic streaming experiments.
}
\label{fig:Fig_5_4}
\end{figure}

In acoustic streaming experiments, the study uses dilute suspensions of 1 $\mu$m fluorescent polystyrene particles (microParticles GmbH, Berlin) dispersed in both Newtonian and viscoelastic (VE) fluids. Deionized water (DI) serves as the Newtonian reference fluid. The VE medium consists of aqueous polyethylene oxide (PEO) solutions with molecular weights of 0.4, 1, and 2 MDa (Sigma-Aldrich, Bangalore, India), prepared at the concentrations listed in Table \ref{tab:5_1}. To ensure complete homogenization, PEO is dissolved in DI water at 25$^\circ$C and stirred continuously for at least 24~h. The fluid viscosities are measured with a rotational rheometer (Anton Paar MCR 72), and their dependence on shear rate is shown in Fig. \ref{fig:Fig_5_4}. For all polymer solutions, the viscosity remains nearly constant up to shear rates of about $400~\textrm{s}^{-1}$, indicating a plateau regime. Only at very high shear rates do the high–molecular weight, high–concentration PEO solutions exhibit a slight decrease in viscosity. Since the shear rates relevant to the acoustic streaming experiments (less than 1$\textrm{s}^{-1}$) are much lower than this range, the solutions effectively exhibit constant viscosity with negligible shear-thinning. This rheological behavior justifies the choice of the Oldroyd-B model for describing the viscoelastic response in the theoretical framework.

In this study, the fluid is subjected to high-frequency longitudinal pressure waves, typically in the MHz range. At these frequencies, the characteristic response of the fluid is governed by timescales distinct from those associated with low-frequency oscillations. To account for this, an approach based on ultrasound spectroscopy \citep{Wu2024,Wu2023,Milliken1990} is employed, which characterizes acoustic absorption and provides a relaxation time directly relevant to the actuation conditions of acoustofluidic systems. Specifically, the study consider the frequency-dependent acoustic absorption coefficient \( \alpha_s(f) \) for PEO solutions, as reported in Refs.~\cite{Wu2024,Wu2023,Milliken1990}, and analyze the relaxation dynamics using the classical Debye relaxation model:
\begin{table}
  \centering
  \setlength{\tabcolsep}{4pt}
  \renewcommand{\arraystretch}{1.15}

  \begin{tabular}{lcccccc}
    \hline
    PEO (MW)
    & $C$ [$\sim$ wt.\%] 
    & $\mu$ [mPa\,s]
    & $\tau$ [$\sim$ $\mu$s]
    & $Dv$
    & $De$ \\[4pt]
    \hline
    1 MDa & 0.05 & 1.20 & $0.43 \pm 0.06$ & 0.35 & $5.20 \pm 0.80$ \\
    1 MDa & 0.10 & 1.69 & $0.14 \pm 0.09$ & 0.90 & $1.75 \pm 1.11$ \\
    1 MDa & 0.50 & 3.19 & $0.80 \pm 0.13$ & 2.59 & $9.80 \pm 1.53$ \\
    2 MDa & 0.18 & 3.66 & $0.12 \pm 0.03$ & 3.12 & $1.40 \pm 0.40$ \\
    2 MDa & 0.35 & 7.12 & $1.65 \pm 0.58$ & 7.01 & $20.04 \pm 7.14$ \\
    2 MDa & 0.52 & 15.31 & $9.10 \pm 2.47$ & 16.21 & $110.45 \pm 30.11$ \\
    0.40 MDa & 0.95 & 15.31 & $0.52 \pm 0.16$ & 16.21 & $6.25 \pm 2.01$ \\
    \hline
  \end{tabular}

  \caption{
  \justifying{Concentration and viscoelastic properties of polyethylene oxide (PEO) solutions.
  The total viscosity is given by $\mu=\mu_s+\mu_p$, where the solvent viscosity is
  $\mu_s=0.89$~mPa\,s (DI water).
  Data compiled from Refs.~\citep{Wu2024,Wu2023,Milliken1990}.}
  }
  \label{tab:5_1}
\end{table}
%

\begin{equation} \label{eq:5_87}
 \alpha_s(f) = \frac{A_1}{1 + \left(\frac{f}{f_r}\right)^2} + B.   
\end{equation}
In this expression, \( f \) denotes the frequency, \( f_r \) is the relaxation frequency, \( A_1 \) represents the relaxation amplitude, and \( B \) accounts for background attenuation from non-relaxation mechanisms. A characteristic decrease in \( \alpha_s \) with increasing frequency indicates the presence of a relaxation process within the observed frequency range. By fitting Eq.~\eqref{eq:5_87} to the experimental absorption coefficient data for the PEO aqueous solutions \citep{Wu2024,Wu2023}, an estimate for the relaxation frequency \( f_r \) is extracted. The corresponding relaxation time is then obtained as \( \tau = 1/f_r \), which agrees well with the values reported in ~\cite{Milliken1990}. This analysis enables a consistent characterization of fluid elasticity across all tested concentrations.

 \subsection{Characterization of acoustic streaming profiles via defocusing particle tracking}

The motion of tracer particles induced by acoustic streaming is observed using an inverted microscope (IX73, Olympus) equipped with a high-speed CCD camera (Phantom), capturing video at 100~frames per second across multiple focal planes. To obtain depth information for the particle positions, a defocusing particle tracking method \citep{Barnkob2015,Barnkob2021} is employed.  

The first step in defocusing particle tracking is to calibrate the system by recording images of tracer particles at known vertical positions or depth, thereby establishing the relationship between particle defocusing and depth. Calibration is performed using the same experimental setup as the measurements. Following ~\cite{Barnkob2015,Barnkob2021}, particles that are either settled at the bottom of the microchannel and attached to the walls are used as reference points. By capturing their defocused images across focal plane, a mapping between the degree of defocusing and the particle's position across channel depth is generated, which enables accurate three-dimensional reconstruction of particle trajectories during acoustic streaming experiments. During calibration, the microscope objective is moved while the tracer particles in the fluid remain stationary. In contrast, during actual experiments, the objective remains fixed while the particles move. To account for this difference, a correction factor is applied to the calibrated depth data. Additionally, since the objective is immersed in air while the particles are suspended in an aqueous dilute polymer solution, a calibration step of $1~\mu\textrm{m}$ corresponds to $1.33~\mu\textrm{m}$ during measurements due to the refractive index mismatch \citep{Barnkob2015,Barnkob2021}.  

Once the calibration is complete using the reference images, tracer particles in the streaming experiments videos are tracked using Tracker 6.1.0, a free video analysis and modeling software, and MATLAB scripts \citep{Barnkob2015,Barnkob2021}. The particle trajectories are then reconstructed in MATLAB and Wolfram Mathematica to obtain the streaming profiles ~\citep{Barnkob2015,Barnkob2021, Karlsen2018, Qiu2021}.

\section{Results and discussion}\label{sec:5_3}

Under a standing acoustic wave imposed along the \(Y\)-direction, acoustic streaming is generated across the cross-section of the microchannel. We perform the analysis on the \(XY\) cross-section (Fig.~\ref{fig:Fig_5_2}a) using the normalized coordinates \( \tilde{X} = X/(W/2) \) and \( \tilde{Y} = Y/(D/2) \), where tildes denote dimensionless variables. We consider dilute polymer solutions exhibiting viscoelastic (VE) behavior with negligible shear-thinning, and model the fluid rheology using the Oldroyd-B constitutive model \citep{oldroyd1950}. This model is characterized by a total viscosity \( \mu = \mu_s + \mu_p \) and a relaxation time \( \tau \), where \( \mu_s \) and \( \mu_p \) denote the solvent and polymer contributions, respectively.

Acoustic streaming is analyzed within the framework of perturbation theory \citep{Bruus2012b,Doinikov2021a,Sujith2024}, accounting for both inner and outer streaming regions associated with distinct characteristic length scales \citep{Bach2018,Lee1990}. A detailed formulation of the theoretical model is presented in  $\S$ \ref{sec:Theory}. The analysis incorporates several relevant time scales, including the acoustic time scale \( t_{ac} = 1/\omega \), the polymer relaxation time \( \tau \), the polymer viscous diffusion time \( t_{d,p} = (D/2)^2/(2\nu_p) \), and the solvent viscous diffusion time \( t_{d,s} = (D/2)^2/(2\nu_s) \), where \( \omega = 2\pi f \), and \( \nu_p \) and \( \nu_s \) are the kinematic viscosities of the polymer and solvent, respectively. The interplay among these time scales is quantified using the Deborah number \( De = \tau/t_{ac} \) and the viscous diffusion number \( Dv = t_{d,s}/t_{d,p} = \nu_p/\nu_s \).

\begin{figure*}\label{fig:Fig_5_5}
\centering
\includegraphics[width=1\columnwidth]
{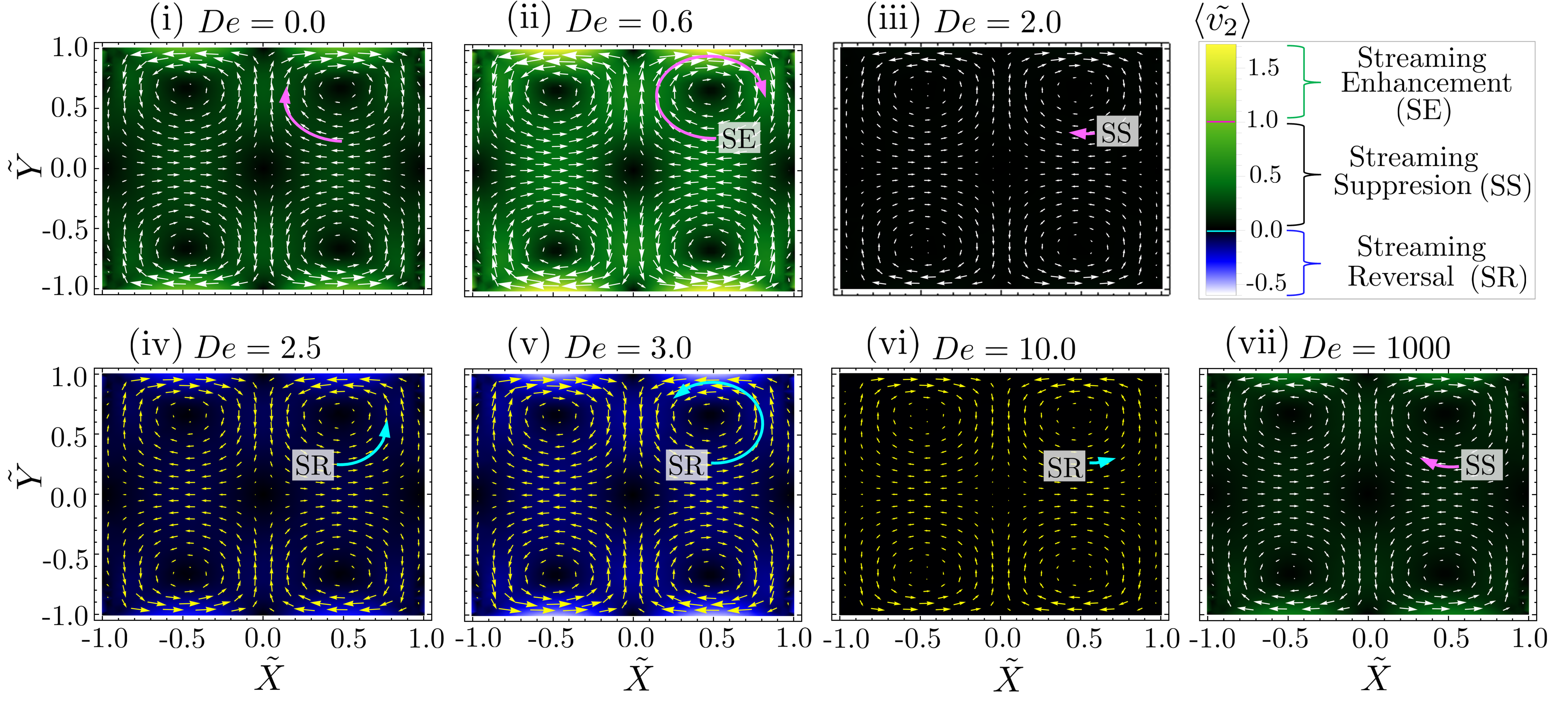}
\caption{\label{fig:Fig_5_5} \justifying Analytical variation of streaming velocity profiles in the channel cross-section (\( XY \) plane) for Deborah numbers \( De = 0 \) to $10^3$ at fixed \( Dv = 20 \). Arrow length and direction represent the relative magnitude and direction of streaming compared to $De=0$ case. (c) Variation of the dimensionless maximum streaming velocity within a quarter of the microchannel cross-section as a function of $De$ at $Dv = 20$, normalized by the Newtonian case $(De=0)$. Labels SE, SS, and SR denote streaming enhancement, suppression, and reversal relative to the purely viscous case ($De = 0$), while $\textrm{SR}^{+}$ and $\textrm{SR}^{-}$ indicate enhancement and suppression in the reversed direction.}
\end{figure*}

\subsection{Effect of viscoelasticity on acoustic streaming}

From the theoretical model (Eqs. \ref{eq:5_73} and \ref{eq:5_74}), the dimensionless streaming velocity \( \langle \tilde{v}_2 \rangle \) response of the VE fluid at \( Dv = 20 \) for varying \( De \) is presented in Fig.~\ref{fig:Fig_5_5}.  Here, \( \langle \tilde{v}_2 \rangle = \langle v_2 \rangle / v_\text{str} \), where \( v_\text{str} = (3/8)(v_1^{d_0})^2/c_0 \) is the Rayleigh streaming velocity, $\langle \cdot \rangle$ denotes the time averaging, \( v_1^{d_0} = A\omega \) is the first-order velocity amplitude, and \( c_0 \) is the speed of sound. 
The variation of dimensionless maximum streaming velocity, \( \langle \tilde{v}_2 \rangle_\text{max}^* = \langle \tilde{v}_2 \rangle_\text{max,VE}/\langle \tilde{v}_2 \rangle_\text{max,N}\) with \( De \), normalized by the Newtonian case (N, $De=0$), is shown in Fig.~\ref{fig:Fig_5_5}. In all cases, four counter-rotating outer streaming rolls are observed. The main findings are: (i) At very low $De$ ($De \leq 1$): viscous effects dominate, and increasing $De$ from 0 (purely viscous case) leads to streaming enhancement (SE). (ii) At low $De$ ($ 1< De \leq 2$): an increase in $De$ gives rise to streaming suppression (SS) as elastic effects become significant and begin to compete with viscous effects. (iii) At moderate $De$ ($2<De\leq4$): there is streaming reversal (SR), and the reversed streaming magnitude increases (SR$^+$) as elasticity becomes more prominent. (iv) At high $De$ ($4<De\leq 10$): the reversed streaming weakens (SR$^-$) as elastic effects decline, leading to streaming suppression. (v) At very high $De$ ($De>10$): the original streaming direction re-emerges with a very weak streaming. These findings suggest that tuning the balance between viscous and elastic effects provides a powerful tool for modulating streaming motion: enabling its enhancement (SE), suppression (SS), or even reversal (SR), offering
a straightforward yet effective strategy for acoustic streaming control. Here, a fundamental question arises: What underlying mechanisms govern the streaming transition? To understand the mechanisms, a distinguish of the two streaming regions: inner streaming in the boundary layer \( \delta_{ve} \), and outer streaming in the bulk is needed. Since \( \delta_{ve} \ll D \) or \( \lambda_0 \), the Stokes parameter \( S_p = \delta_{ve}/D \ll 1 \). This scale separation allows decomposition of the field into a boundary region (``\(\delta\)'') and a bulk region (``\(d\)'') (see Fig.~\ref{fig:Fig_5_2}b).

\begin{figure*}
\includegraphics[width=0.98\columnwidth]{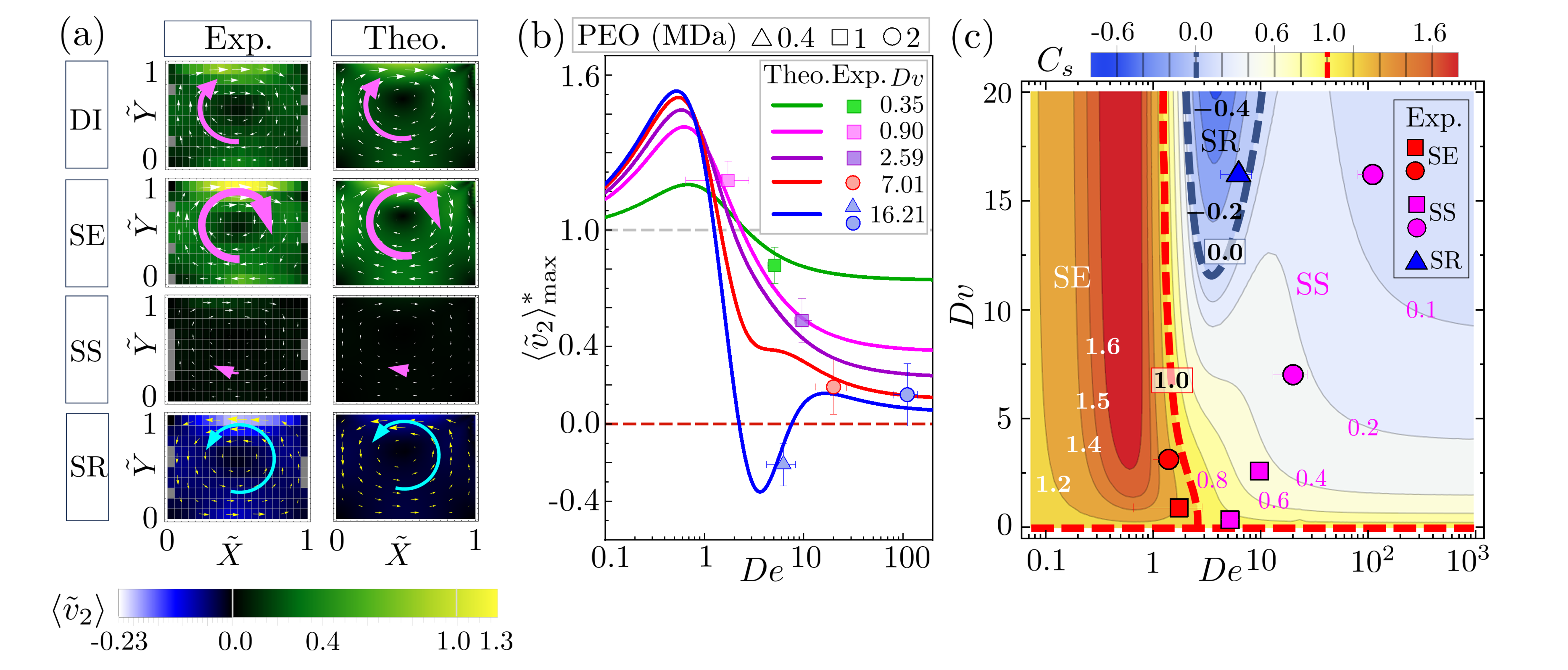}
\centering
\caption{\label{fig:Fig_5_6}  \justifying (a) Comparison of theoretical and experimental  dimensionless streaming velocity $\langle \tilde{v}_2 \rangle$ profiles across one quarter of the channel cross-section, illustrating streaming enhancement (SE, PEO 1 MDa, $Dv=0.90$, $De=1.75$), streaming suppression (SS, PEO 2 MDa, $Dv=16.21$, $De=110.45$), and streaming reversal (SR, PEO 0.4 MDa, $Dv=16.21$, $De=6.25$), compared with deionized water (DI, $Dv=0$, $De=0$). Arrows indicate streaming direction. (b) Theoretical variation of the dimensionless maximum streaming velocity \( \langle \tilde{v}_2 \rangle_{\textrm{max}}^* \) with \( De \) for PEO solutions, with experimental data shown as symbols. (c) Analytical prediction of the streaming coefficient $C_s$ as a function of $De$ and $Dv$; symbols denote experimental data. Red ($C_s=1$) and blue ($C_s=0$) dashed lines mark the SE–SS and SS–SR transitions, respectively.
}
\end{figure*}\label{fig:Fig_5_6}

The inner and outer acoustic streaming fields (Fig.~\ref{fig:Fig_5_2}) are derived analytically using a second-order perturbation expansion of the continuity and momentum equations \citep{Sujith2024}. The first-order acoustic fields exhibit harmonic time dependence of the form \( e^{i\omega t} \), and the velocity field is decomposed using the Helmholtz representation \citep{Bach2018} as \( \boldsymbol{v}_1 = \boldsymbol{v}_1^d + \boldsymbol{v}_1^\delta \), where the irrotational component is given by \( \boldsymbol{v}_1^d(x,t) = v_1^{d0} \cos(k_x x) e^{i\omega t} \), with \( k_x = 2\pi/\lambda_0 \) denoting the acoustic wavenumber. The time-averaged second-order velocity field is similarly decomposed as \( \langle \boldsymbol{v}_2 \rangle = \langle \boldsymbol{v}_2^d \rangle + \langle \boldsymbol{v}_2^\delta \rangle \). Solution of the boundary-layer (\(\delta\)) equations yields the inner streaming field and an effective slip-velocity boundary condition at the walls \citep{Lee1990}, which drives the bulk (outer) streaming flow \citep{Rayleigh1884}. The outer (\(d\)) streaming solution for parallel plates is extended to rectangular microchannels with aspect ratio \( \beta = D/W \) using a Fourier series expansion \citep{Muller2013}. The complete mathematical derivation is provided in $\S$~\ref{sec:Theory}, from which the time-averaged second-order streaming velocity in the bulk is obtained,

\begin{equation}\label{eq:5_88}
\langle \tilde{\boldsymbol{v}}_{2}(\tilde{X},\tilde{Y})\rangle = C_s \langle \tilde{\boldsymbol{v}}_{2}(\tilde{X},\tilde{Y})\rangle_N 
\end{equation}
%

\noindent In Eq.~\ref{eq:5_88}, \( \langle \tilde{\boldsymbol{v}}_{2} \rangle_N \) represents the Rayleigh streaming velocity for a Newtonian fluid in a rectangular microchannel \citep{Muller2013}, while \( C_s \) is the \textit{streaming coefficient} introduced to account for viscoelastic effects. The theoretical predictions are validated against experiments performed using \(1\,\mu\mathrm{m}\) fluorescent polystyrene tracer particles suspended in polyethylene oxide (PEO) solutions with molecular weights of 0.4, 1, and 2~MDa (see Fig.~\ref{fig:Fig_5_6} a–c). These measurements provide experimental evidence of streaming enhancement (SE), suppression (SS), and reversal (SR) in viscoelastic fluids confined within microchannels. The spatial variation of the streaming velocity over one quarter of the channel cross-section is quantified experimentally using the defocusing particle tracking technique \citep{Barnkob2015,Barnkob2021} and compared with theoretical predictions obtained from Eq.~\ref{eq:5_88}, as shown in Fig.~\ref{fig:Fig_5_6}(a). Here, SS and SE occur over a wide range of fluid properties. In contrast, SR appears only within a narrow parameter window and is highly sensitive to experimental conditions and fluid preparation, making it challenging. In addition, the predicted dependence of the normalized maximum streaming velocity, \( \langle \tilde{v}_2 \rangle_{\mathrm{max}}^* \), exhibits good agreement with the experimental data (see Fig.~\ref{fig:Fig_5_6}b).

\subsection{Streaming Coefficient: regime characterization}

 To characterize the streaming transitions in viscoelastic fluids shown in Fig.~\ref{fig:Fig_5_5}, the streaming coefficient \( C_s \) is introduced to quantify deviations from classical Rayleigh streaming in Newtonian fluids (see Eq.~\ref{eq:5_88}). Streaming is enhanced for \( C_s > 1 \), suppressed for \( 0 \leq C_s \leq 1 \), and reversed for \( C_s < 0 \). In the derivation of streaming (see Eq. \ref{eq:5_50}), the streaming coefficient is obtained as
\begin{equation}\label{eq:5_050}
 C_s=\frac{2 \rho_0 c_0^2 I_1+2 De  Dv \mu_s^* I_2+ 2 De^2 I_3}{3(1+De^2)\chi^2 (\chi^2+\xi^2)^2 \rho_0 c_0^2  \mu_s^*(1+Dv)} 
\end{equation}

\noindent Here $\rho_0$ denotes the fluid density, $\mu_s^*=\mu_s/\mu_f$, $\mu_f$ is the viscosity of the base Newtonian fluid (deionized water), $I_1,I_2$ and $I_3$ are functions of $Dv$ and $De$, as given in Eqs. (\ref{eq:5_51})-(\ref{eq:5_53}).

The variation of \( C_s \) with \( De \) and \( Dv \) is presented in Fig.~\ref{fig:Fig_5_6}(c). A regime of streaming enhancement (SE), characterized by \( C_s > 1 \), is observed at low \( De \) over the entire range of \( Dv \), where viscous effects dominate. At higher values of \( Dv \) (\( Dv > 12 \)), increasing \( De \) leads to a transition from SE to a narrow streaming suppression (SS) regime (\( 0 \leq C_s \leq 1 \)), followed by streaming reversal (SR) (\( C_s < 0 \)), and subsequently a return to SS at high \( De \). In contrast, for \( Dv < 12 \), the SE regime transitions into a broadened SS regime without the occurrence of reversal. The experimental data shown in Fig.~\ref{fig:Fig_5_6}(c) fall within the same regimes predicted by the theoretical model, confirming quantitative consistency between theory and experiments. These trends identify \( Dv > 12 \) as a necessary condition and \( 2 < De < 10 \) as a sufficient condition for the onset of SR, underscoring the coupled roles of viscous and elastic effects. To further elucidate the physical mechanisms underlying these transitions, the relative contributions of Reynolds stress (RES) and viscoelastic stress (VES) within the acoustic boundary layer are analyzed.

\subsection{Streaming patterns and effect of Reynolds and viscoelastic stresses }

 Outer streaming is driven by inner streaming ~\citep{Rayleigh1884,Schlichting1932}; thus, the transitions observed in Fig.~\ref{fig:Fig_5_5} and the regimes in Fig.~\ref{fig:Fig_5_6}(c) stem from changes within the boundary layer. The model shows that the time-averaged slip velocity \( \langle \tilde{v}_2^{\delta} \rangle_s \), which governs outer streaming, depends on fluid viscoelasticity. This slip velocity arises from nonlinearities in the boundary layer and modulates the bulk flow. In Newtonian fluids, inner streaming is driven by RES \citep{Rayleigh1884,Muller2012}. In VE fluids, both RES and VES contribute to the acoustic body force \( f_{ac} \), which drives the streaming and is given by 
\begin{multline}\label{eq:5_90}
f_{ac}= \boldsymbol{\nabla} \cdot \rho_0 \left\langle\boldsymbol{v}_1^\delta \boldsymbol{v}_1^\delta+ \boldsymbol{v}_1^\delta \boldsymbol{v}_1^d+\boldsymbol{v}_1^d \boldsymbol{v}_1^\delta\right\rangle+ \boldsymbol{\nabla} \cdot \tau \Bigl\{ \left\langle\boldsymbol{v}_1^\delta \cdot \boldsymbol{\nabla} \boldsymbol{\sigma}_1\right\rangle  \\  -
\left\langle\left(\boldsymbol{\nabla} \boldsymbol{v}_1^\delta\right)^T \cdot \boldsymbol{\sigma}_1\right\rangle- \left\langle\boldsymbol{\sigma}_1 \cdot \boldsymbol{\nabla} \boldsymbol{v}_1^\delta\right. \left.\right\rangle \Bigr\}
\end{multline}
\noindent Here, \(\boldsymbol{\sigma}_1\) denotes the first-order viscoelastic stress. The right-hand side of Eq.~(\ref{eq:5_90}) contains the divergence of RES and VES. In VE fluids, RES comprises viscous (\(\textrm{RES},v\)) and elastic (\(\textrm{RES},ve\)) components. These stresses collectively drive the flow, which is opposed by viscosity, and this induces a momentum flux gradient, forming streaming rolls. Elasticity alters dissipation, modifying these gradients and enabling flow enhancement, suppression, or reversal (see Fig.~\ref{fig:Fig_5_5}). To quantify this, the streaming slip velocity is considered as \(\langle \tilde{v}_2^{\delta} \rangle_s = S_{\text{RES},v} + S_{\text{RES},ve} + S_{\text{VES}}\).

\begin{figure}
    \centering
    \includegraphics[width=0.75\linewidth]{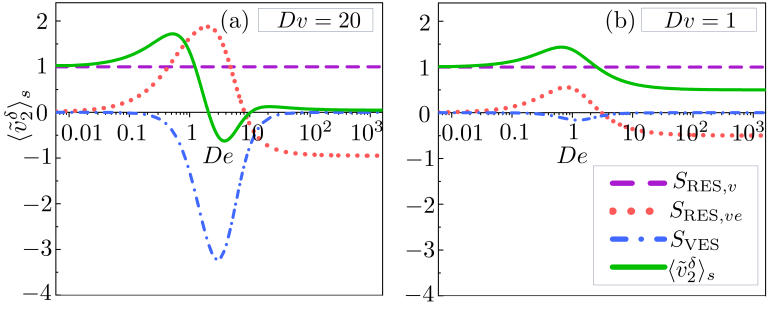} 
    \caption{\justifying Variation of the streaming slip velocity $\langle \tilde{v}_2^{\delta} \rangle_s$ and the contributions of the viscous Reynolds stress (\( S_{\text{RES},v} \)), viscoelastic Reynolds stress (\( S_{\text{RES},ve} \)), and viscoelastic stresses (\( S_{\text{VES}} \)) with \( De \) for (a) $Dv=20$ and (b) $Dv=1$.}
    \label{fig:Fig_5_7}
\end{figure}

The variation of \( \langle \tilde{v}_2^{\delta} \rangle_s \) along with contributions from viscous RES (\( S_{\text{RES},v} \)), viscoelastic RES (\( S_{\text{RES},ve} \)), and VES (\( S_{\text{VES}} \)) as a function of \( De \) for various \( Dv \) is shown in Fig.~\ref{fig:Fig_5_7}. The complete expressions for these contributions are provided in $\S$~\ref{sec:2_04}. At \( Dv = 20 \), \( \langle \tilde{v}_2^{\delta} \rangle_s \) initially increases with \( De \), then decreases to zero, later reverses, and finally re-emerges in the original direction, consistent with transitions across SE, SS, SR, and SS (see Fig.~\ref{fig:Fig_5_5}). \( S_{\text{RES},v} \) remains constant and positive, \( S_{\text{RES},ve} \) grows to a maximum then declines and turns negative near \( De \approx 10 \). \( S_{\text{VES}} \) is always negative, peaking in magnitude before declining with increasing $De$. At high \( Dv \) (Fig.~\ref{fig:Fig_5_7}a), the following findings are made:   (i) at very low \( De \), \( |S_{\text{RES},ve}| > |S_{\text{VES}}| \), leading to SE. (ii) at low \( De \), \( |S_{\text{VES}}| \approx |S_{\text{RES},v}| + |S_{\text{RES},ve}| \), resulting in SS. (iii) at moderate \( De \), \( |S_{\text{VES}}| > |S_{\text{RES},v}| + |S_{\text{RES},ve}| \), causing SR.  (iv) at high \( De \), \( |S_{\text{VES}}| + |S_{\text{RES},ve}| \approx |S_{\text{RES},v}| \), leading to suppression in the reversed streaming. (v) at very high \( De \), \( |S_{\text{VES}}| \) becomes negligible, 
 highly suppressed streaming in the original direction is observed.  In contrast, at low \( Dv \) (\( Dv = 1 \)), \( S_{\text{VES}} \) is negligible throughout and \( \langle \tilde{v}_2^{\delta} \rangle_s \) stays positive across \( De \), showing only SE and SS (Fig.~\ref{fig:Fig_5_7}b). Although \( S_{\text{RES},ve} \) turns negative for \( De > 2 \), the condition \( |S_{\text{RES},ve}| + |S_{\text{VES}}| < |S_{\text{RES},v}| \) prevents reversal, therefore SS persists beyond \( De > 2 \). Notably, both RES and VES arise from nonlinear interactions within the boundary layer, governed by the 
 \( v_1^{\delta} \) (see Eq.~\ref{eq:5_90}), which can be attributed to viscoelastic shear waves \citep{Stokes_2009, Ferry1942, Pascal1989, Ortn2020}.

\subsection{Viscoelastic shear Waves: streaming patterns}

The interaction between oscillatory perturbations along the \( x \)-direction (\( v_1^d \)) and the top and bottom walls generates VE shear waves \citep{Stokes_2009, Ferry1942, Pascal1989, Ortn2020} that propagate from the walls into the bulk (see inset in Fig.~\ref{fig:fIG_5_8}a). These waves, emerging from viscous dissipation and linked to acoustic energy loss, are characterized by an attenuation length (or boundary layer thickness) \( \delta_{ve} \) and a shear wavelength \( \lambda_{ve} \). For \( S_p = \delta_{ve}/D \ll 1 \), the system is in the surface-loading regime~\citep{Ortn2020}, where shear waves decay before reaching the opposite wall, the $x$-component of the shear wave is
\begin{equation}\label{eq:5_91}
   {v}_{1,x}^\delta=-v_1^{d_{0}} \cos \left(k_x x\right)  e^{-\chi y/ \delta_{f}} e^{i (\omega t-\xi y/ \delta_{f})}
\end{equation}
The \( y \)-component of the shear wave is negligible as \( \delta_f k_x \ll 1 \), where \( \delta_f = \sqrt{2\mu_f/\rho_0 \omega} \) is the attenuation length in the base fluid. From Eq.~(\ref{eq:5_91}), \( \boldsymbol{v}_1^{\delta} = \hat{\boldsymbol{v}}_1^{\delta} e^{i \omega t} \), where \( \hat{\boldsymbol{v}}_1^{\delta} \) is the time-independent amplitude. The variation of \( \tilde{v}_1^{\delta} \ (= \hat{v}_{1,x}^{\delta}/v_1^{d_0}) \) along \( y \) at various \( De \) is plotted in Fig.~\ref{fig:fIG_5_8}(a). For \( De = 0 \), the wave is overdamped; increasing \( De \) yields underdamped behavior with varying \( \delta_{ve} \) and \( \lambda_{ve} \).

We consider a complex propagation constant \( P = AF + i\,PF \), with \( AF = 1/\delta_{ve} = \chi/\delta_f \) quantifying amplitude decay, and \( PF = 2\pi/\lambda_{ve} = \xi/\delta_f \) capturing phase variation. The ratio \( \alpha = AF/PF \) along with \( De \) and \( Dv \) characterizes VE shear wave behavior (see Fig. ~\ref{fig:fIG_5_8}b). For \( \alpha = 1 \), shear waves are overdamped, yielding  streaming similar to viscous fluids. However, the underdamped waves (\( \alpha < 1 \)) correspond to streaming transitions. At high \( Dv \), SE occurs for \( 0.4 < \alpha < 1 \), SS for \( 0.28 < \alpha < 0.4 \), and SR for \( \alpha \leq 0.28 \). At low \( Dv \), \( \alpha > 0.28 \) for all \( De \), indicating absence of SR. Hence, a critical VE parameter is identified, \( \alpha_c = 0.28 \) that marks the onset of SR. However, how do these waves induce reversal at this threshold? To answer this, the energy dynamics of VE shear waves is analyzed.

Acoustic energy dissipation is key to generating momentum flux gradients that drive streaming \citep{LIGHTHILL1978}. Like longitudinal waves, VE shear waves also transport energy. To understand their role, complex shear impedance \( Z = C + iR \) is analyzed, where \( R \sim G'' \) (energy dissipation or loss modulus) and \( C \sim G' \) (energy storage and release or storage modulus), linked to the complex shear modulus \( G = G' + iG'' \). The \( G' \) and \( G'' \)  \citep{ferry1980} are 
\begin{equation}
    G'=\frac{\rho_0 v_s^2 (1-\alpha^2)}{(1+\alpha^2)^2}, \quad G''=\frac{2\rho_0 v_s^2 \alpha }{(1+\alpha^2)^2.}
\end{equation}

\begin{figure*}
\includegraphics[width=1.0\columnwidth]{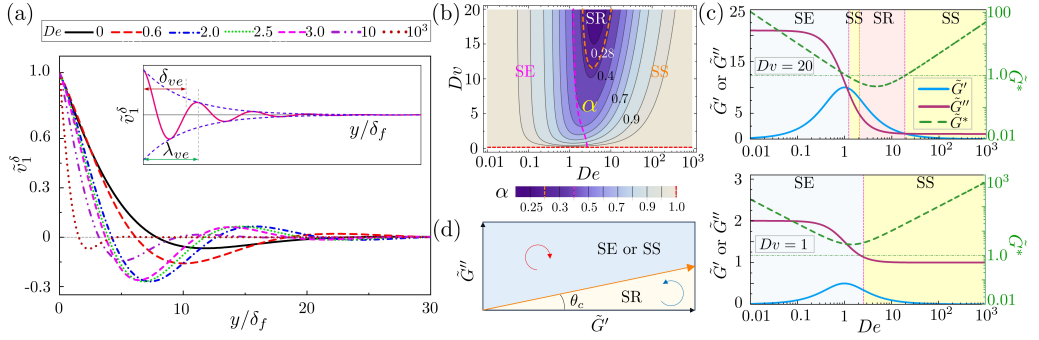}
\caption{\label{fig:fIG_5_8} \justifying (a) Viscoelastic shear wave profiles along $y/\delta_f$ for varying $De$ at $Dv = 20$ and $\tilde{x} = 0$. Inset: schematic showing attenuation length $\delta_{ve}$ and shear wavelength $\lambda_{ve}$. (b) Variation of $\alpha \  (=\lambda_{ve}/2 \pi \delta_{ve})$ with $De$ and $Dv$. (c) Variation of storage modulus $\tilde{G}'$, loss modulus $\tilde{G}''$, and their ratio $\tilde{G}^* = \tilde{G}''/\tilde{G}'$ with $De$ for $Dv = 20$ and $Dv = 1$. (d) Phase lag $\theta = \tan^{-1}(\tilde{G}^*)$ showing critical lag $\theta_c$ marking the onset of SR. Magenta and orange dashed lines indicate SE–SS and SS–SR transitions. 
`}
\label{fig:fIG_5_8}
\end{figure*}

\noindent Here, \( v_s = \omega \lambda_{ve} / 2\pi \) is shear wave velocity. For \( \alpha = 1 \), fluid behaves Newtonian. The moduli is normalized using loss modulus of base fluid (DI water): \( \tilde{G}' = G'/G''_f \), \( \tilde{G}'' = G''/G''_f \), and define loss tangent as \( \tan(\theta) = \tilde{G}''/\tilde{G}'=\tilde{G}^* \).  A high $\tan(\theta)$ indicates rapid energy dissipation, while a low value signifies efficient storage and release.

The variation of $\tilde{G}'$, $\tilde{G}''$, and $\tilde{G}^*$ with $De$ at high $Dv$ is presented in Fig.~\ref{fig:fIG_5_8}(c). At very low $De$, the response is strongly dissipative: $\tilde{G}'' \gg \tilde{G}'$, $\tilde{G}^* \gg 1$, and $\tilde{G}''$ remains constant, indicating a viscous-dominated regime with rapid shear wave relaxation ($0.4 < \alpha < 1$). This leads to SE via strong momentum flux gradients (see Fig.~\ref{fig:Fig_5_7}). As $De$ increases, $\tilde{G}'$ peaks at $De=1$ then reduces due to reduced polymeric contributions at high relaxation times \citep{Doinikov2021a, Sujith2024}, while $\tilde{G}''$ decreases monotonically, reaching a minimum near $De=10$. At $De=1$, $\tilde{G}' \approx \tilde{G}''$ and $\tilde{G}^* \approx 1$, marking the SE--SS transition. Further increase in elasticity yields $\tilde{G}'' < \tilde{G}'$ and $\tilde{G}^* < 1$, reducing dissipation and altering momentum flux. At moderate and high $De$ with $\alpha < \alpha_c$, $\tilde{G}' \gg \tilde{G}''$, VE shear waves reverse flow by modifying time-averaged stresses and momentum gradients. However, at very high $De$ ($De \gg 1$), $\alpha \approx 1$ and $\tilde{G}'' > \tilde{G}'$, the original flow direction resumes. At low $Dv$ ($Dv=1$), SE transitions to SS with increasing $De$, but reversal is absent as $\tilde{G}' < \tilde{G}''$ throughout. In contrast, high $Dv$ amplifies $\tilde{G}'$, enabling reversal once a threshold is exceeded.

To quantify the interplay of moduli, the phase lag $\theta = \tan^{-1}(\tilde{G}^*)$ is shown in Fig.~\ref{fig:fIG_5_8}(e).
A purely viscous response or viscous shear wave corresponds to \(\theta = \pi/2\), while an ideal elastic response or elastic shear wave gives \(\theta = 0\). At $\alpha_c$, $\tan(\theta_c)=\tilde{G}^* \approx 0.60$ gives $\theta_c \approx 31.03^\circ$ (Fig.~\ref{fig:fIG_5_8}d). Streaming remains enhanced or suppressed for $\theta > \theta_c$, but reverses for $\theta < \theta_c$. This highlights the need for a sufficiently high storage modulus relative to the loss modulus for flow reversal, linking viscoelasticity to energy dissipation and momentum flux modulation.

\section{Conclusion}

In conclusion, a theoretical framework is introduced for the first time to describe acoustic streaming in viscoelastic fluids confined within rectangular microchannels. The study demonstrates that acoustic streaming in a rectangular microchannel can be enhanced (SE), suppressed (SS), or reversed (SR) through controlled variation of fluid viscoelasticity. The streaming behavior is primarily governed by the Deborah number (\( De \)) and the viscous diffusion number (\( Dv \)), which together regulate the balance between Reynolds and viscoelastic stresses via energy dissipation in shear waves. The streaming transitions are characterized using \( \tilde{G}^* \), defined as the ratio of the loss to storage modulus of the shear wave. At high \( Dv \), streaming enhancement occurs for \( \tilde{G}^* \gg 1 \), suppression for \( \tilde{G}^* \approx 1 \), and reversal for \( \tilde{G}^* < 1 \), whereas at low \( Dv \), only the transition from enhancement to suppression is observed. These viscoelasticity-driven streaming transitions offer new opportunities for controlled bioparticle manipulation and fluid pumping or mixing in microfluidic systems.

\bibliographystyle{jfm}
\bibliography{jfm}

\end{document}